\documentstyle[12pt,epsfig]{article}

\newcounter{multieqs}

\newcommand{\bq}{\begin{equation}}
\newcommand{\fq}{\end{equation}}
\newcommand{\bqr}{\begin{eqnarray}}
\newcommand{\fqr}{\end{eqnarray}}
\newcommand{\non}{\nonumber \\}

\newcommand{\rf}[1]{(\ref{#1})}

\def\x{\vec{x}}

\newcommand{\half}{ {\textstyle \frac{1}{2}}}
\newcommand{\ts}{\textstyle}

 \def\E{{\cal E}} 

  \def\cI{{\cal I}}
\def\I{{\cal I}}

 \def\O{{\cal O}}

\def\N{{\cal N}}

\def\pr{\partial}

\begin{document}

\thispagestyle{empty}
\setcounter{page}{0}

\begin{flushright}
\begin{tabular}{l}
ANL-HEP-PR-00-055 \\ 
MIT-CTP-2986 \\
hep-th/0005192
\end{tabular}
\end{flushright}

\vspace{10mm}

\begin{center}

{\Large \bf Dual Expansions of $\N=4$ Super Yang-Mills}\\ 
\vskip .12in
{\Large \bf Theory via IIB Superstring Theory}\\

\vspace{10mm}

{\bf Gordon Chalmers} \\[3mm]
{\em Argonne National Laboratory \\
High Energy Physics Division \\
9700 South Cass Avenue \\
Argonne, IL  60439-4815 } \\  

\vspace{10mm} 

{\bf Johanna Erdmenger} \\[3mm] 
{\em Massachusetts Institute of Technology \\
Center for Theoretical Physics \\ 
77 Massachusetts Avenue
 \\ 
Cambridge, MA  02139-4307 } \\ 

\vspace{10mm}

{\bf Abstract}  
\end{center}

We examine the dual correspondence between holographic IIB superstring 
theory and $\N=4$ super Yang-Mills theory at finite values of the coupling 
constants. In particular we analyze a field theory strong-coupling expansion 
which is the S-dual of the planar expansion.  This expansion arises 
naturally as the AdS/CFT dual of the IIB superstring scattering amplitudes  
given a genus truncation property due to modular invariance.
The space-time structure of the contributions to the field theory four-point 
correlation functions obtained from the IIB scattering elements is 
investigated  in the example of the product of four conserved stress 
tensors, and is expressed as an infinite sum of field theory triangle integrals.  
The OPE structure of these contributions to the stress tensor 
four-point function is analyzed and shown not to give rise to any poles.  
Quantization of the string in the background of a five-form field strength 
is performed through a covariantized background field approach, 
and relations to the $\N=4$ topological string are found.

\vfill 
\line(6,0){220} \vskip .04in
e-mail: chalmers@pcl9.hep.anl.gov, jke@mitlns.mit.edu \hfill

\break

\newpage 
\setcounter{footnote}{0}

\section{Introduction}  

Holographic string scattering within the conjectured duality between 
$\N=4$ super Yang-Mills theory and IIB superstring theory 
\cite{Maldacena:1998re,Gubser:1998bc, Witten:1998qj} has led to many interesting 
results within the low-energy supergravity approximation of string theory.  
According to the Maldacena conjecture, the string theory parameters
$\alpha'$ and $g_s$, the inverse string tension and the string coupling,
are related to the field theory parameters $N$ and $\lambda$, 
the number of colors and the 't Hooft coupling, by
\bqr  
 {L^4 \over \alpha'^2} = \lambda=g^2_{\rm YM}N \, ,  \qquad \qquad 
 g_s ={1\over 4\pi} \, g^2_{\rm YM} \ , 
\label{vardefin}
\fqr 
where $L$ is the AdS radius. The supergravity approximation to string theory
is valid if the string coupling is small and if the inverse string tension is
much smaller than the square of the AdS radius. In the dual field theory this 
regime pertains to both $N$ and $\lambda$ very large with $g_{\rm YM}$ small.  
In this approximation a variety of $\N=4$ super Yang-Mills theory correlation 
functions have been calculated and properties of the strong coupling limit of 
the gauge theory found using the AdS/CFT correspondence (an exhaustive list of 
references may be found in \cite{Aharony:1999ti}).  Non-trivial renormalization 
properties of the correlation functions, as well as the correspondence between 
field theory operators and string states, provide evidence for the AdS/CFT 
correspondence.  For correlation functions satisfying a non-renormalization 
theorem, such as two- and three-point functions involving conserved currents or 
scalar chiral primary operators, these results have been used successfully 
to test the AdS/CFT correspondence by finding agreement between the strong coupling 
AdS results and weak coupling perturbative field theory calculations 
\cite{Freedman:1999tz,Muck,Chalmers:1999xr,Lee:1998bx,D'Hoker:1999tz, Skiba:1999im}.  
Further non-trivial tests of the correspondence in the supergravity approximation 
are provided by non-renormalization theorems for extremal $n$-point correlation 
functions with $n \geq 4$ and by non-trivial renormalization properties of
related correlators \cite{frex,SM,MJ,MJDE,HWex}.

However holographic string scattering, as formulated 
in \cite{Witten:1998qj}, provides a generating functional also for finite values 
of the coupling constant.  By virtue of the AdS/CFT duality, 
this formulation provides additional constraints on both string 
and field theory given that the AdS/CFT duality holds at finite values of the 
couplings.  The OPE structure of $\N=4$ super Yang-Mills theory is directly related 
to the string scattering, and also to the truncation of the string theory to its 
massless modes in a string-inspired regulated supergravity context as discussed 
in \cite{Chalmers:2000zg}.  

A natural extension of the supergravity limit investigations towards the intermediate 
coupling regime is to relax the conditions of the supergravity low-energy limit 
and to consider string contributions to the correlation functions. This amounts 
to investigating the consequences of the duality in its ``strong'' form, i.e. to 
assume its validity for {\it any} value of $N$ and $\lambda$, and to analyzing 
its consequences both for the boundary field theory and for the superstring itself.  
A first step is the consideration of corrections to the massless mode 
scattering of supergravity in an expansion in the inverse string tension 
and in the string coupling in the holographic context \cite{Banks:1998nr}, as 
opposed to a direct path integral quantization in the five-form background. This weakly 
coupled parameter region of string theory with $\alpha'$ and $g_s$ small corresponds 
on the field theory side to the parameter region where
\begin{equation} 
\label{weak}
\lambda = g_{\rm YM}^2 N \;\;\;\; {\rm large} \, ,\;\;\;\; 
\;\;\;\;g_{\rm YM}^2 = \frac{\lambda}{N}
\; \;\;\; {\rm small} \, . 
\end{equation}
For the stress tensor four-point function, the first non-trivial $\alpha'$
correction was considered  in \cite{Banks:1998nr} 
with regards to instantons within the AdS/CFT 
correspondence. On the string theory side, this 
correction is given by \cite{GG}
\begin{equation}
S^{IIB} = \frac{1}{{\alpha'}^4} \, \int \! {\rm d}^{10} x \;
\sqrt{-g} \,  {\alpha'}^3 \, e^{-\phi/2} \, f^0(\tau , \bar \tau)
t^8 t^8 R^4  \, , \label{BG}
\end{equation}
where $R$ is the curvature tensor contracted in the way well-known from the 
tree and genus one scattering amplitude in IIB string theory.  The tensor $t^8$ 
and its contraction with $R^4$ will be given in more detail below. $f^0(\tau, 
\bar\tau)$ is a non-holomorphic Eisenstein series, $E_{3/2}(\tau,\bar\tau)$, 
whose form is strongly constrained by the $SL(2,Z)$ invariance of IIB superstring 
graviton scattering.  In the quantum supergravity limit where string modes  
are absent there is a piece of this structure that remains upon specification 
of the string-related regulator of the supergravity.   Furthermore in the classical 
limit of supergravity, there is a $SL(2,R)$ symmetry which has implications
for $\N=4$ super Yang-Mills
at infinite $N$ and $\lambda$, as discussed discussed in detail 
in \cite{Intriligator:1999ig,Intriligator:1999ff}.  To the derivative order given 
by \rf{BG}, the low-energy scattering of gravitons (and of the other fields of the 
string) receives tree and genus one contributions from the string perturbation 
expansion in $g_s$, as well as non-perturbative contributions from D instantons.  
Within AdS/CFT, the argument $\tau$ in (\ref{BG}) is related to the Yang-Mills 
variables by
\bqr  
\tau= {\theta\over 2\pi} + i {4\pi \over g_{\rm YM}^2} \, . 
\fqr 
$f^0(\tau , \bar \tau)$ has the correct coupling constant dependence to agree with 
field-theory instanton calculations.  

In \cite{Banks:1998nr} it was shown that $ AdS_5 \times S_5$ remains a solution
of the string theory equations of motion in the presence of the term (\ref{BG}) 
to all orders in $g_s$. This is essentially due to the symmetry properties of 
the $t^8$ tensors which ensure that only the Weyl tensor part of the curvature 
tensors contributes to (\ref{BG}).  Since the space $ AdS_5 \times S_5$ is 
conformally flat, its Weyl tensor vanishes.  The space-time dependence of the 
four-point function contribution obtained from the four-dilaton term in the low-energy 
S-matrix in (\ref{BG}) was calculated in \cite{Brodie}, where it was shown for
holographic dilaton scattering that its short-distance singularities are at most 
logarithmic.

On the string theory side, several extensions of (\ref{BG}) to higher
orders in $\alpha'$ have been discussed.  These include terms involving
the five-form field strength $F_5$ \cite{Berkovits:1998ex}, a direct 
$SL(2,Z)$ covariantization of the IIB tree-level graviton scattering involving 
derivatives acting on the $R^4$ structure \cite{Russo:1998fi,Russo:1998vt}, 
as well perturbation theory based on a manifest $SL(2,Z)$ invariant theory 
\cite{Chalmers:1999ap}.  The $\Box^2 R^4$ next to leading term was analyzed 
at genus two in \cite{Green:2000pu}.  Supersymmetry constraints on the string 
expansion to lowest order were investigated in \cite{Sethi}.  In all of these cases 
$S$-duality or modular invariance play a crucial role in determining the 
coupling dependence in terms of Eisenstein series (See also \cite{Obers:2000um, 
Obers:1999fb}).  For graviton scattering, an extension of the S-matrix 
that is compatible with perturbative string theory and the known results in 
maximal supergravity, as well as with unitarity of the massless sector, 
has been given in \cite{Chalmers:2000zg}.  

Via the AdS/CFT correspondence, all of these string theory scattering amplitudes 
give rise to contributions to the stress tensor four-point function at strong 't Hooft coupling.  
The $\lambda$ and $N$ dependence of these contributions corresponds 
to a resummation of the perturbative series as well as to a 't Hooft expansion 
around an instanton background.  The fact that the coupling dependence of the 
string theory amplitudes is strongly constrained by modular invariance implies 
on the field theory side that the 't Hooft large $N$ expansion, 
for which
\begin{equation}
\lambda = g_{\rm YM}^2 N \; \;\; \;{\rm small \; and \; fixed} \, , \;\;\;\;
\;\;\;\;
N \rightarrow \infty \, , \;\; 
\label{tHooft}
\end{equation}
has only a finite number of perturbative terms in $\lambda$ at each order 
in $N$, considered for the instantonic terms in \cite{Gopakumar}.  Of course, the 
derivation of this result requires the assumption that 
the large $\lambda$ expression obtained from string theory may be analytically 
continued to the weak coupling regime determined by the 't Hooft limit (\ref{tHooft}).  
The instanton sector non-renormalization theorem following from this assumption, 
where ``non-renormalization'' means that there are only finitely many perturbative 
terms for a given order in $1/N$, is consistent with the results of 
\cite{Bianchi,Dorey1,Dorey2}, where instanton contributions are calculated to 
leading order in $\alpha'$.  We refer the reader to \cite{Belitsky:2000ws} for 
a comprehensive review.  

In this paper we consider the extension of the IIB superstring 
S-matrix to all orders in $\alpha'$ and $g_s$ given in \cite{Chalmers:2000zg}. 
By virtue of the AdS/CFT correspondence we examine its consequences
for $\N=4$  super Yang-Mills theory.  In the superstring theory, there  
are, for example, higher derivative terms in the covariantized S-matrix 
which are generated by 
\bqr  
S_{4-pt}^{\rm IIB} = \frac{1}{\alpha'}
\sum_{k=0}^\infty \int d^{10}x \sqrt{g} ~ 
\alpha'{}^k \,  f_k(\tau,\bar\tau) \Box^k R^4 \ \, . 
\label{sform1}
\fqr 
Here $\Box^{k}R^4$ is a shorthand notation for derivatives acting on some of the 
curvature tensors.  The $\Box^{k}R^4$ terms have the same supersymmetry weight as 
the $R^4 H^{4k}$ amplitudes of the $\N=4$ topological string \cite{Berkovits:1995vy}.  
As discussed in \cite{Chalmers:2000zg}, the functions $f_k$ in (\ref{sform1}) are 
again severely constrained by modular invariance, which strongly suggests in particular 
that at order $k$ in $\alpha'$, there are only finitely many perturbative corrections 
up to a maximum genus $g_{\rm max}={1\over 2} (k+2)$ for $k$ even or $g_{\rm max}= 
{1\over 2}(k+1)$ for $k$ odd.  We refer to this behaviour as the {\it genus truncation 
property}. 

Using the strong form of the AdS/CFT correspondence, we calculate the contributions 
to the stress tensor four point function arising from (\ref{sform1}) and from further 
polynomial terms in the string scattering elements.  
We derive the $\lambda$ and $N$ dependence of these contributions 
from the functions $f_k$ using the relations (\ref{vardefin}).  A further essential 
ingredient of our analysis, motivated by the importance of modular invariance in this 
context, is to reorganize the series for the coupling dependence into terms involving 
the S-dual of the 't Hooft coupling $\lambda$.  Under $S$-duality we have $g_{\rm YM} 
\rightarrow 1/g_{\rm YM}$ such that
\begin{equation}
\lambda= g^2_{YM} N\, \;\;\;\;\rightarrow \,\;\;\;\; 
\lambda_D = \frac{N}{g^2_{\rm YM}} = 
\frac{N^2}{\lambda} \, .
\end{equation}
Furthermore the S-dual of the original 't Hooft limit (\ref{tHooft})
is given by
\begin{equation}
\lambda_D = \frac{N^2}{\lambda} \,\;\;\;\; {\rm small \; and \; fixed} 
\, , \;\;\;\;\;\;\;\;N \rightarrow
\infty \, .  
\label{dualplanar}
\end{equation}
From the point of view of the original 't Hooft coupling $\lambda$ this is clearly a
strong coupling limit since both $\lambda$ as well as the Yang-Mills coupling
$g_{\rm YM}$ are very large in this limit\footnote{For a first consideration of an expansion
in $\lambda_D$ see \cite{Eguchi:1998rt}.}.

We find that in this dual 't Hooft limit the perturbation series in $\lambda_D$
arising from the modular functions $f_k$ contains just the terms corresponding
to the highest genus contribution for each $\alpha'$ on the string theory side, 
given that the genus truncation property holds. This leads us to the result that 
{\it given the strong form of the AdS/CFT correspondence, the genus truncation 
property is equivalent to the convergence of the strong coupling perturbative 
expansion of $\N = 4$ SYM in $\lambda_D$ in the dual 't Hooft limit.}  

Since $g_{\rm YM}$ is no longer small in the dual 't Hooft limit we are
outside the parameter region given by (\ref{weak}).  We do not prove the 
convergence of the S-dual to the planar series here.  However the original 
proof of the convergence of the planar series in $\lambda$ \cite{'tHooft:1982tz}, 
according to which the planar expansion has a finite radius of convergence, 
in combination with the strong implications of S-duality, provides a strong 
argument in favour of the convergence of the $\lambda_D$ series in the dual 
't Hooft limit.

Furthermore, we determine the spatial dependence of the contributions to the 
stress tensor four-point function arising from (\ref{sform1}), which correspond to
contact diagrams for the graviton scattering, and 
investigate  their short-distance limit when two of the four point approach 
each other. In particular, we show that (\ref{sform1}) does not contribute 
any leading terms to the OPE.  This implies that there are no poles, which 
is consistent with the instanton nature of (\ref{sform1}), and also with Ward 
identities relating the four-point function to the three-point function for which 
a non-renormalization theorem holds.  Another consequence is that these contact contributions 
to the stress tensor four-point function do not contain any term which factors 
into two two-point functions (in contrast to the exchange contributions, 
which also contribute non-trivially to the Ward identities).  

The outline of this work is as follows.  In section 2 we examine the 
generating function of correlation functions given by the holographic string 
scattering of gravitons dual to the composite stress-tensor operators.  
We analyze the coupling constant structure and re-organize the planar 
expansion into its S-dual relative.  In section 3 we examine the latter 
series and convergence with regards to the modular properties of the 
string scattering.  In section 4 we derive a generating functional as 
a one-parameter integral representation that resums an infinite class of 
known contributions of the string scattering (with similar resummations 
in integral form).  We also examine the unitarity structure of the holographic 
scattering with regards to the correlation functions in the dual gauge theory.  
The space-time structure of the correlation functions are examined in section 
5, and using conformal inversions, 
the holographic contact diagrams are expressed 
as triangle one-loop integral functions. The tensor structure is obtained by extracting
appropriate derivatives.  We also consider 
related box diagrams relevant for contact diagrams as well as for
massless supergravity exchange diagrams.  
The pole structure of the finite coupling constant dual 
correlators is examined and shown to be sub-leading away from the infinite 
coupling limit when only the contact diagrams contribute.  
In section 7 we conclude and discuss related avenues.  

\section{$\N=4$ Correlation functions from IIB Quantum S-matrix} 
\setcounter{equation}{0}

In this section we examine the $\alpha'$ and $g_s$ expansion of IIB superstring 
theory in the background of the $AdS_5\times S^5$ space with regards to deriving 
the super Yang-Mills theory correlators and their coupling dependence.  We begin by 
discussing terms in the $\alpha'$ expansion which lead to contact diagrams for 
the four-graviton scattering.  Then we discuss 
terms in the expansion which lead to exchange diagrams. 
Subsequently we show how Yang-Mills correlators are obtained by virtue of the AdS/CFT 
correspondence and discuss their coupling dependence in detail, in particular 
the implications of the genus truncation property.  

As discussed in \cite{Chalmers:2000zg}, the low-energy derivative expansion 
of the IIB superstring S-matrix in ten-dimensions for energies such that 
$s<4/\alpha'$ involves contributions of the form
\bqr  
S_{4-pt}^{\rm IIB} = \sum_{k=0}^\infty {1\over\alpha'} \int d^{10}x \sqrt{g} ~ 
 f_k(\tau,\bar\tau) (\alpha'\Box)^k R^4 \ 
\label{sform}
\fqr 
in Einstein frame and where $f_k(\tau,\bar\tau)$ are invariant 
non-holomorphic functions under fractional linear transformations 
of the IIB string coupling constant $\tau=\chi+ie^{-\phi}$, which we 
label as $\tau=\tau_1+i\tau_2$.  The individual terms in \rf{sform} are 
\bqr 
\Box^k R^4 \rightarrow t^8_{\mu_1\nu_1\ldots \mu_4\nu_4} t^8_{{\bar\mu}_1 
{\bar\nu}_1\ldots {\bar\mu}_4{\bar\nu}_4} \Box^k\Bigl( R^{\mu_1\nu_1{\bar\mu}_1{\bar\nu}_1}  
 R^{\mu_2\nu_2{\bar\mu}_2{\bar\nu}_2} \Bigr) R^{\mu_3\nu_3{\bar\mu}_3{\bar\nu}_3}
R^{\mu_4\nu_4{\bar\mu}_4{\bar\nu}_4} \ , 
\fqr 
together with mixed derivatives on the curvature tensors (at tree-level in 
momentum space the tensor has the form $s^k +t^k+u^k$).  The expansion in \rf{sform} 
pertains to the four-point graviton scattering amplitude; higher $n$-point 
amplitudes involve further curvature tensors (related to additional polarization 
vectors).  The generic $R^4$ structure has not been explicitly shown to be the 
unique tensor at higher-genus, although there is evidence based on multi-loop IIB 
supergravity scattering \cite{Bern:1998ug} as well as in multi-genus superstring 
scattering; further symmetry structures of the indices constrain the possible forms, as 
discussed below.  We consider the generic $R^4$ tensor structure to be a consequence of 
$\N=8$ supersymmetry.  In momentum space, $\Box^k$ represents the symmetrized factor 
\bqr  
\Box^k \rightarrow s^k + t^k + u^k \ ,  
\fqr 
with the invariants 
\bqr  
s=(k_1+k_2)^2 \qquad t=(k_1+k_3)^2, \qquad u=(k_2+k_3)^2 
\fqr 
at tree-level, and more general (unknown) combinations $s^{k_1} t^{k_2} 
u^{k_3}$ (with $k_1+k_2+k_3=k$) at higher-genus.  Note that $s+t+u=0$ and 
$s^3+t^3+u^3=3stu$ via momentum conservation. 

Furthermore, the R$\otimes$R five-form effects coupled with the $R^4$ 
have the expansion  
\bqr  
S^{F}_{4-pt} = \sum_{k=1}^\infty {1\over\alpha'} \int d^{10}x \sqrt{g} ~ g_k(\tau,
\bar\tau) \alpha'^{2k} F_5^{4k} R^4 \ . 
\label{fiveform}
\fqr 
Consistent classical propagation of the string requires the beta function 
conditions 
\bqr  
\langle R_{\mu\nu} \rangle = e^{\phi} 
\langle F_{5,\mu}{}^{\mu_1\ldots\mu_4} F_{5,\nu\mu_1 
\ldots\mu_4} \rangle \ , 
\fqr 
and we take this to hold at higher order (potential corrections at higher order 
in $\alpha'$, i.e. at four loops, potentially produce an $R^4$ counterterm which 
vanishes in the background as required by stability).  

The tree-level coefficient in \rf{fiveform} in string frame has the coupling 
constant   
\bqr 
g_k^{\rm tree} = \tau_2^{-5k+2} 
\label{gktree}
\fqr 
with successive $\tau_2^{-2}$ at higher genus; the conjecture in 
\cite{Berkovits:1998ex,Bsem} generates the form of \rf{fiveform} at 
genus zero and genus $k-1$ for a given $k$.  Lastly, a multi-point graviton 
scattering amplitude generates terms in the quantum generating functional 
of the S-matrix of the form 
\bqr  
S^{4+n;k} = \sum_{i_n=1}^\infty {1\over\alpha'} \int d^{10}x \sqrt{g} ~ h^{(n)}_{i_n} 
(\tau,\bar\tau) (\alpha')^{2k+n} F_5^{4k} R^{4+n} \ , 
\label{multipoint}
\fqr 
where a tensor contraction of the curvature tensors is implied and may 
be found by performing amplitude calculations in the IIB superstring.  The 
analysis of the coupling constant structure of these terms in \rf{multipoint} 
and \rf{fiveform} is similar to that in \rf{sform}.

In the background field approach, the substitution of the $AdS_5 
\times S^5$ metric into \rf{multipoint} leads to a contraction 
of four $R_{\mu\nu}$ terms in the expansion, and the question 
arises as to how many independent tensors this term may generate.  
The Weyl symmetries and those of the four-graviton scattering 
permit in general three independent tensors, and requiring stability 
of the background implies that the contribution of the higher-derivative 
term in \rf{multipoint} gives rise to the same $R^4$ term as before 
but with a coupling constant structure associated with the higher 
derivatives.   The form has an overall factor of 
\bqr  
{1\over L^{2n+4k}} \alpha'^{2k+n} = {1\over \lambda^{k+n/2}} \ ,  
\fqr 
in addition to the relative higher string coupling constants associated 
with the genus corrections in $h^{(n)}(\tau,\bar\tau)$.  

The form in \rf{fiveform} and in \rf{sform} may be found by 
performing explicit S-matrix calculations in IIB superstring 
theory.  In \rf{fiveform} a background value $\langle F_5\rangle$  
and $g_{\mu\nu}^{AdS}$ ($\langle R^{\mu\nu}{}_{\rho\gamma}\rangle$) 
generates additional contributions in the covariantized scattering 
to those obtained from that in \rf{sform} within the AdS/CFT 
correspondence.  Note that in the on-shell quantum corrected effective 
action there is no requirement to have an off-shell IIB description 
of the action for the self-dual five-form.  

In addition to \rf{fiveform} and \rf{sform} which lead to contact bulk 
four-point interactions for gravitons on the anti-de Sitter space, there 
are also exchanges of the massless modes at tree-level.  They contribute 
a term  of the form 
\bqr 
S_{m=0}^{4-pt,IIB} = {1\over\alpha'} 
 \int d^{10}x \sqrt{g} ~ {1\over (\alpha'\Box)^3} R^4 \ , 
\label{masslessexchange}
\fqr 
to the quantum corrected on-shell action found from the S-matrix.  
Here the inverted $\Box$'s are arranged so that in momentum space the factor 
$stu$ is found.  In the holographic context these exchange contributions have 
been analyzed in a number of works and correspond to the infinite $N$ and 
$\lambda$ results.  There are further non-analytic terms in the S-matrix 
related to the thresholds of the massless modes and unitarity, which may be 
constructed iteratively following the analysis in \cite{Chalmers:2000zg} from 
the $SL(2,Z)$ construction of matrix elements in IIB superstring theory.  

We now investigate the coupling dependence of these S-matrix contributions.  
The functional form in \rf{sform} represents the polynomial terms that arise 
from integrating out the massive modes at genus $g\geq 0$ and the massless ones at 
$g\geq 1$ in the genus expansion (in the string field theory this arises in a 
well-specified regulator).  An ansatz for the functional form of 
$f_k(\tau,\bar\tau)$ using Eisenstein series has been proposed in 
\cite{Chalmers:2000zg}.  The general perturbative structure of $f_k(\tau,\bar\tau)$ 
required by the dilaton dependence of IIB superstring perturbation theory is 
\bqr 
f_k(\tau,\bar\tau) = a_0^{(k)} \tau_2^{{3\over 2}+{k\over 2}} 
+ a_1^{(k)} \tau_2^{-{1\over 2}+{k\over 2}} + a_2^{(k)} 
 \tau_2^{-{5\over 2}+{k\over 2}} + \ldots \ , 
\label{form}
\fqr 
which is a decreasing power series in the string coupling 
constant $\tau_2^2$.  In \rf{form}, the index $k$ labels the order in $\alpha'$ (twice 
the number of derivatives with respect to the eight derivative $R^4$ term) and 
the subscript on $a_i$ the contributing genus order.  In \cite{Chalmers:2000zg} 
it was conjectured that this series truncates, i.e. for a given $k$ the S-matrix 
receives perturbative corrections up to a maximum genus $g_{\rm max}={1\over 2} 
(k+2)$ ($k$ even) and $g_{\rm max}= {1\over 2}(k+1)$ ($k$ odd).  The corresponding
set of terms in the low-energy expansion is listed in Table 1.  This genus 
truncation is in agreement with the modular structure of S-duality, together 
with the perturbative dependence in uncompactified IIB string theory together 
with further consistency conditions of the S-matrix.  Furthermore, it is in 
agreement with the genus truncation of the conjectures for the amplitudes 
$R^4 H^{4g-4}$ in \cite{Berkovits:1998ex}.  The non-analytic terms in the 
superstring S-matrix contribute orders in the derivative expansion below 
the starred entries in Table 1.  The leading primitive divergences of the 
supergravity are polynomials in momenta, with the non-analytic terms non-leading.  
In $\N=4$ super Yang-Mills theory these terms will therefore contribute only 
at sub-leading in $N$.   

\begin{table}  
\caption{
Example contributions to $\Box^k R^4$ arising in string 
perturbation theory at genus $g$.  The asterisk denotes the top 
genus contributions.} 
\begin{center}
\begin{tabular}{|l|l|l|l|l|l|r|} \hline 
{\em } &  $g=0$ & $g=1$ & $g=2$ & $g=3$ & $g=4$ & $\ldots$ \\ \hline 
$R^4$      & $\surd$ & $\surd^\star$ & & & & \\ \hline  
$\Box R^4$ &         &         & & & & \\ \hline  
$\Box^2 R^4$ & $\surd$ &   & $\surd^\star$ & & &  \\ \hline 
$\Box^3 R^4$ & $\surd$ & $\surd$ & $\surd$ & & & \\ \hline 
$\Box^4 R^4$ & $\surd$ & $\surd$ & $\surd$ & $\surd^\star$ & & \\ \hline 
$\Box^5 R^4$ & $\surd$ & $\surd$ & $\surd$ & $\surd$ & & \\ \hline 
$\Box^6 R^4$ & $\surd$ & $\surd$ & $\surd$ & $\surd$ & $\surd^\star$ & \\ \hline
$\cdots$ & & & & & & \\ \hline     
\end{tabular} 
\end{center} 
\end{table} 

From the expansion in \rf{sform} we may derive quantum corrected 
equations of motion and generate the finite $\lambda$ and $N$ terms 
in the boundary $\N=4$ gauge theory; in the following we consider the 
contributions from \rf{sform} for simplicity. The remaining terms in 
\rf{multipoint} may be analyzed in a similar way, and we shall comment 
on their contributions to the analysis when appropriate. 

For obtaining the contribution of (\ref{sform}) to the stress tensor
four-point function, we consider fluctuations around the background
geometry $AdS_5 \times S_5$ and vary four times with respect to the
five-dimensional fluctuation around $AdS_5$. Then we restrict to
$AdS_5 \times S_5$. The five-sphere contributes just a constant factor
which we omit in the subsequent.
 
For the $AdS_5$ geometry we use the conventions and notations of 
\cite{Freedman:1999tz}. In particular, for the square of the five-dimensional 
line element we have $ds^2={1\over z_0^2}(dz_0^2+d{\vec z}^2)$, where
$z_0\geq 0$ and ${\vec z}\in R^d$. The background solution for the $AdS_5$ 
compactification is 
\bqr 
F_{\mu_1\ldots\mu_5}= {1\over L} \epsilon_{\mu_1\ldots\mu_5} \qquad 
R_{\mu_1\ldots\mu_4} = {1\over L^2}\bigl( g_{\mu_1\mu_3}g_{\mu_2\mu_4} - 
g_{\mu_1\mu_4}g_{\mu_2\mu_3}\bigr)  \ .
\fqr 
The curvature of the AdS space, $L$, is set to unity unless otherwise 
stated. The form in \rf{vardefin} translates tree exchange of the modes 
of the superstring and the higher genus orders in the holographic string 
theory on $AdS_5 \times S^5$ to $\lambda$ and $N^2$ effects, 
respectively, in the $\N=4$ gauge theory.  

\begin{figure}
\begin{center}
\epsfig{file=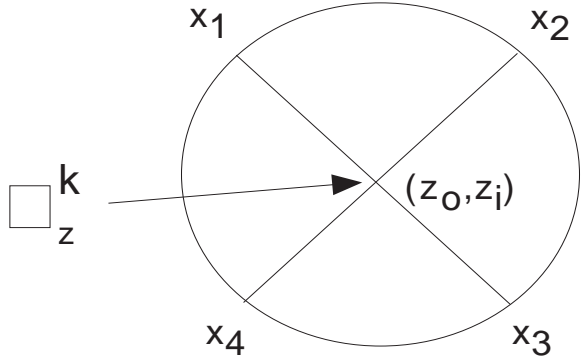,height=5cm,width=8cm} 
\end{center} 
\caption{Holographic effective four-point contact diagram found from the 
covariantized fourth variation of the on-shell IIB string scattering 
element.  The $\Box$ denotes the action $\Box^k R^4$.  } 
\end{figure} 

In this way we obtain for the effective four-point vertex 
\bqr  
V_g^{(k),\{\mu_i\nu_i\} }(z_0,{\vec z}) \equiv a_g^{(k)} \tau_2^{3/2+k/2-2g} 
 \alpha^{'k+3}~ \prod_{j=1}^4 \bigl\{ {\delta\over
\delta {\hat g}(z_0,{\vec z})} \bigr\} \Bigl[ \sqrt{g} ~ \Box^k R^4 \Bigr]  
\vert_{g_{AdS_5\times S^5}} \ , 
\label{effvertex} 
\fqr 
\bqr  
= a_g^{(k)} \tau_2^{3/2+k/2-2g} \alpha^{'k+3}~ {\tilde V}_g^{(k), 
 \{\mu_i\nu_i\} }(z_0, {\vec z}) \ , 
\fqr 
where ${\hat g}$ is the fluctuation around the background geometry and 
$a_g^{(k)}$ is a numerical constant related to the genus $g$ and order 
$\alpha'^k$ term in the expansion at small $\alpha'$.  The vertex in 
\rf{effvertex} generates the holographic Feynman-Witten diagram in Figure 1 
and is dual to the contribution to the correlation of four stress-tensors 
by virtue of
\bqr  
\langle \prod_{j=1}^4 T^{a_j b_j}({\vec x}_i) \rangle = N^2 \sum_{k=0}^\infty 
\sum_{g=0}^{g_k} \lambda^{-k/2} \Bigl({\lambda\over N}\Bigr)^{-3/2+2g}  
a_g^{(k),\{a_j b_j\}}({\vec x}_i) 
\label{genexpand}
\fqr 
\bqr 
a_g^{(k),\{a_i b_i\}} ({\vec x}_i) = a_g^{(k)} ~ \int {dz_0 d^d{\vec z}\over 
z_0^{d+1}}  ~ 
 \Bigl[ \prod_{i=1}^4 G^{\{a_ib_i\}}{}_{\{\mu_i\nu_i\}} 
 ({\vec x}_i; z_0,{\vec z}) \Bigr] ~ {\tilde V}_g^{(k),\{\mu_i\nu_i\}} 
 (z_0,{\vec z}) \ ,  
\label{axfunction}
\fqr 
for $\N=4$, $d=4$ ($AdS_{d+1}$). Here $G^{ab}{}_{\mu \nu} 
({\vec x}_i; z_0,{\vec z})$ is the appropriate 
bulk-boundary kernel. The space-time dependence of (\ref{genexpand}) is 
discussed in detail in Section \ref{OPEsec} below.

We shall suppress the external graviton indices associated with 
the bulk-boundary kernel, 
\bqr  
a_k^{(g)}({\vec x}_i) = a_k^{(g),\{a_j b_j\}} ({\vec x}_i) \ , 
\fqr 
when it is clear from the text.  Furthermore  we define
\bqr  
\langle \prod_{j=1}^4 T^{a_j b_j}({\vec x}_i) \rangle = N^2~ \sum_{k=1}^\infty 
\sum_{g=0}^{g_k} f_g^{(k),\{a_jb_j\}}({\vec x}_j) \ , 
\label{fnotation}
\fqr 
where $g$ labels the genus order and $(k)$ the derivative order ($k=0$ 
corresponds to eight derivatives) in the genus expansion.  
We define the the functions $f$ without 
an overall factor of $N^2$
and note that  they express both the coupling dependence as well as the 
space-time structure.  
The space-time indices in \rf{fnotation} 
will be suppresssed  in the subsequent discussion of the coupling dependence.  
Furthermore we define 
\bqr  
f^{(k)} = \sum_{g=0}^{g_k} f_g^{(k)} \ , \label{sum1}
\fqr 
and 
\bqr 
f_g = \sum_{k_{\rm min}}^{\infty} f^{(k)}_g \, ,  \quad 
\quad k_{\rm min} = 2 (g_{\rm max} -1) \, ,
 \label{sum2}
\fqr   
where $g_{\rm max}$ is the maximum even genus in agreement with the genus truncation property
as defined in the introduction.
Low-energy holographic scattering in the bulk string corresponds in the dual 
gauge theory to a power series in the coupling around infinite $\lambda$. 
Integrating out modes in the string and fixing energy scales of the scattering 
at lower values (hence probing larger distances) does not correspond to energy 
scales in the dual field theory but rather the region in coupling space accessible 
in the approximation around infinite coupling; this translates the Wilsonian 
sense of energy scales to the boundary data and we must integrate out to more 
orders in the bulk to access finite coupling in the dual gauge theory.  

The massless graviton exchange given by \rf{masslessexchange} represents 
the supergravity approximation and has the coupling constant structure to be 
the leading $N$ and $\lambda$ contribution to the four-point correlation function.  
Its general form for the massless field exchange has been found in a number 
of recent papers \cite{Liu:1999ty,Chalmers:1999wu,D'Hoker:1999mz,D'Hoker:1999jc, 
D'Hoker:1999pj}, and generically corresponds to coordinate space box diagrams 
in the boundary field theory with rational function coefficients in the position 
variables.  Contact diagrams arising from integrating massive string exchange diagrams 
produces similar box integrals\footnote{the unitarity properties of which 
were analyzed in \cite{Chalmers:1999xj} as a function of $\lambda$ and large $N$.}, 
although in this case the results are found both by performing an $\alpha'$ expansion 
at higher genus as well as at genus zero as far as the massive modes are concerned.  
We find that the field theory box structure appears generically to all orders in 
$\lambda$ and $N$ in this approach, including the integration over the massless 
propagating modes, as illustrated in Figure 2. 

We shall step through the expansion in \rf{genexpand} to demonstrate 
the coupling structure from the point of view of IIB superstring theory and 
the regrouping of the terms in the S-dual variable.  The general form of the 
large $\lambda$ and $N$ expanded correlation functions is found by using the 
covariantized S-matrix in \rf{sform} to leading order in $N^2$ through the 
$\alpha'$ expansion.  Using the $\tau_2^2$ expansion of \rf{sform}, we list 
explicitly the functional dependence arising from $\Box^k R^4$ for $k=0$ 
through $k=4$ in the $\N=4$ gauge theory:  From the $R^4$ term
we have, using the notation (\ref{sum1}),  
\bqr  
f^{(0)}({\vec x}_i) &=& {1\over\lambda^{3/2}}  
 \Bigl[ a_0^{(0)}({\vec x}_i) + a_1^{(0)}({\vec x}_i) \Bigl({\lambda\over N}\Bigr)^2 
 \Bigr] \ , 
\fqr  
with $a_0^{(0)}$ the genus zero and $a_1^{(1)}$ the genus one terms of the 
gauge theory.  The $\Box R^4$ vanishes on-shell due to momentum conservation, 
and the next ones are, from $\Box^2 R^4$, 
\bqr   
f^{(2)}({\vec x}_i) &=& {1\over\lambda^{5/2}} \Bigl[ a_0^{(2)}({\vec x}_i) + 
 a_2^{(2)}({\vec x}_i) \Bigl({\lambda\over N}\Bigr)^4 \Bigr] \ , 
\fqr  
from $\Box^3 R^4$, 
\bqr     
f^{(3)}({\vec x}_i) &=& {1\over\lambda^{3}} \Bigl[ a_0^{(3)}({\vec x}_i) + 
 a_1^{(3)}({\vec x}_i) \Bigl({\lambda\over N}\Bigr)^2 + a_2^{(3)}({\vec x}_i) 
\Bigl({\lambda\over N}\Bigr)^4 \Bigr] \ ,
\fqr 
and from $\Box^4 R^4$, 
\bqr  
f^{(4)}({\vec x}_i) = {1\over\lambda^{7/2}} \Bigl[ a_0^{(4)}({\vec x}_i) + 
 a_1^{(4)}({\vec x}_i) \Bigl({\lambda\over N}\Bigr)^2 + a_2^{(4)}({\vec x}_i) 
\Bigl({\lambda\over N}\Bigr)^4 + a_3^{(4)}({\vec x}_i)  
 \Bigl({\lambda\over N}\Bigr)^6 \Bigr] \ .  
\label{fseriesexample}
\fqr 
The general dependence from \rf{sform} is 
\bqr  
f^{(k)}({\vec x}_i) = \lambda^{-3/2-k/2} ~\sum_{g=1}^{g_{\rm max}(k)} 
 a_g^{(k)}({\vec x}_i) \bigl({\lambda\over N}\bigr)^{2g} \ . 
\label{fseries}
\fqr 
In the above we assumed the genus truncation property to hold, but we 
shall analyze possible higher order contributions in the next section.  

\begin{figure}
\begin{center}
\epsfig{file=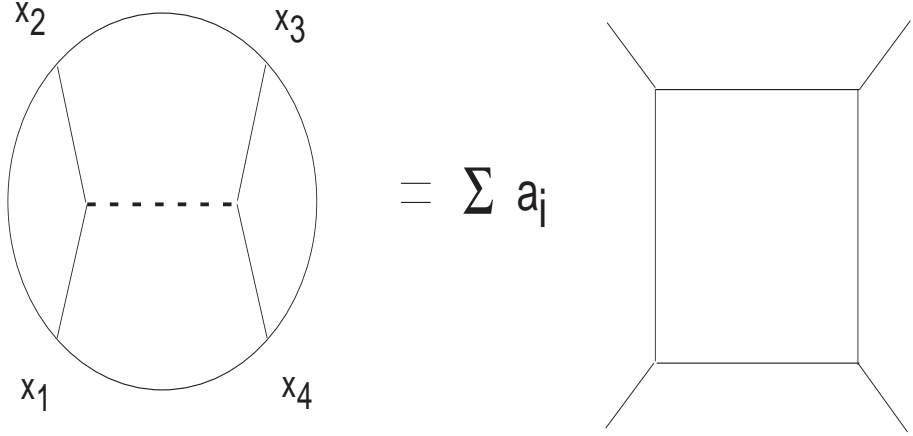,height=6cm,width=12cm} 
\end{center} 
\caption{Massless field exchange produces linear combinations of box diagrams 
with rational function coefficients $a_j$.} 
\end{figure} 

We now rearrange the series in \rf{fseries} in a manner so that the $N^2$ 
dependence is manifest.  The leading $N$ dependence is captured by the first 
term in every series listed in \rf{fseries}. We use the notation (\ref{sum2}).
In the classical string theory 
limit we have, expanding in the string scale   
\bqr 
f_0({\vec x}_i)= {1\over N^3} \Bigl({N^2\over \lambda}\Bigr)^{3/2} 
\Bigl[ a_0^{(0)}({\vec x}_i) + {1\over\lambda} a_0^{(2)}({\vec x}_i)  + 
 {1\over \lambda^{3/2}} a_0^{(3)}({\vec x}_i) +
{1\over\lambda^2} a_0^{(4)}({\vec x}_i)  + \ldots \Bigr] \ . 
\label{genuszero}
\fqr  
Sub-leading $N^2$-suppressed contributions are taken from the 
non-leading functions in \rf{fseries}.  Each sub-leading series 
contains an infinite number of terms and are explicitly, from genus one, 
\bqr 
f_1({\vec x}_i)= {1\over N} \bigl( {\lambda\over N^2}\bigr)^{1/2} \Bigl[
a_1^{(0)}({\vec x}_i) + {1\over\lambda^{3/2}} a_1^{(3)}({\vec x}_i)  + {1\over \lambda^2}
a_1^{(4)}({\vec x}_i) + {1\over\lambda^{5/2}} a_1^{(5)}({\vec x}_i)  + \ldots \Bigr] \ ,
\fqr 
from genus two, 
\bqr 
f_2({\vec x}_i)= {1\over N} \bigl( {\lambda\over N^2}\bigr)^{3/2} \Bigl[
a_2^{(2)}({\vec x}_i) + {1\over\lambda^{1/2}} a_2^{(3)}({\vec x}_i)  + {1\over \lambda}
a_2^{(4)}({\vec x}_i) + {1\over\lambda^{3/2}} a_2^{(5)}({\vec x}_i)  + \ldots \Bigr] \ ,
\fqr 
and from genus three, 
\bqr 
f_3({\vec x}_i)= {1\over N} \bigl( {\lambda\over N^2}\bigr)^{5/2} \Bigl[
a_3^{(4)}({\vec x}_i) + {1\over\lambda^{1/2}} a_3^{(5)}({\vec x}_i)  + {1\over \lambda}
a_3^{(6)}({\vec x}_i) + {1\over\lambda^{3/2}} a_3^{(7)}({\vec x}_i)  + \ldots \Bigr] \ .
\label{Nexpansion}
\fqr 
The relative factors to the leading ones in the above series are 
\bqr  
{1\over \lambda^{1/2}} = {1\over N} \Bigl( {N^2\over\lambda}\Bigr)^{1/2} \ , 
\label{factor}
\fqr 
and for fixed $N^2/\lambda$ are $1/N$ suppressed as $N$ becomes large. 
The higher genus contributions have a factor of $N^2$ relative to 
the first series (i.e. classical superstring theory) in \rf{genuszero} after 
using the expansion in the dual coupling via \rf{factor}; the 
classical string contribution is sub-leading in $N$ in the expansion of 
$\lambda/N^2$.  Specific contributions are listed in \rf{genuszero}-\rf{Nexpansion}.  
The perturbative series for the correlation function at large coupling 
$\lambda$ (and $N$) is encoded in the equation, 
\bqr  
f_j({\vec x}_i) = \lambda^{-1/2} 
 \Bigl[{\lambda\over N^2}\Bigr]^j~ \sum_{k=2j-2}^\infty \lambda^{-k/2} 
~ a_j^{(k)}({\vec x}_i) \ ,
\label{leadingkwise}
\fqr 
where $j$ denotes the genus order and $k$ the $\alpha'$ order in the 
low-energy expansion of the superstring graviton scattering.  

In taking the dual 't Hooft limit in \rf{dualplanar} the perturbative 
series is re-organized and different terms in $f_j({\vec x}_j)$ contribute.
The strong coupling limit  
\bqr 
{\lambda\over N^2} ={\rm fixed}, \qquad N\rightarrow\infty  
\label{duallimit} \ ,
\fqr  
isolates from the sum of the functions $f_j({\vec x}_i)$, or 
$f^{(k)}({\vec x}_i)$, the leading dependence               
\bqr  
\langle \prod_{j=1}^4 T^{a_j b_j}({\vec x}_i) \rangle\vert_{S-dual} = 
N^2 \lambda^{-1/2} ~\sum_{g=1}^\infty
\Bigl({\lambda\over N^2}\Bigr)^{g}  a_g^{(2g-2),\{\mu_j\nu_j\}}({\vec x}_j) \ .  
\label{leadingduallimit} 
\fqr  
The remainder to \rf{leadingduallimit} has terms of order ${1\over\lambda^{1/2}} 
= {1\over N} ({N^2\over\lambda})^{1/2}$ higher and are individually suppressed 
in this limit.  A similar expansion in $1/N$ is available by collecting the 
terms in \rf{leadingkwise} at lower orders in $\lambda$.  The series in 
\rf{leadingduallimit} does not contain terms in tree-level string scattering 
in the background, and as such does not have a classical supergravity limit. 

The important aspect of the series in \rf{leadingduallimit} in the string 
context is that for every $\alpha'$ it just involves the maximum even genus 
contribution in agreement with the {\it genus truncation property} of 
\cite{Chalmers:2000zg}, which is defined after \rf{sform1} of the present 
paper.  On the Yang-Mills side, \rf{leadingduallimit} is an expansion in 
the S-dual of the 't Hooft coupling to leading order in $1/N$.  We see 
that a field theory proof of the convergence of the expansion in 
\rf{leadingduallimit} would imply the genus truncation property in string 
theory.  Vice-versa, a string theory proof of the genus truncation property 
would imply the convergence of the field theory strong coupling expansion.  
We note that 't Hooft's original proof of the convergence of the large $N$ 
planar expansion holds for a finite radius of convergence; together with 
the S-duality between the original planar expansion and the one considered 
here this provides strong evidence for the possibility of a field theory 
proof for the convergence of \rf{leadingduallimit}.   
The limit in \rf{duallimit} 
describes the dual planar limit of $\N=4$ gauge theory, and hence the 
monopole and dyonic dual description to the planar expansion in $1/N$.  

From the weak coupling point of view, the limit in \rf{duallimit} does 
not have a microscopic formulation because the gauge theory does not 
generate graphs perturbatively with the dependence in \rf{dualplanar}.  
The general diagram, for $SU(N)$, of genus $g$ with Lagrangian normalization 
\bqr 
{\cal L} =  {N\over 2 \lambda} \int d^dx ~ \Bigl( -{1\over 2} 
{\rm Tr} F^{\mu\nu} F_{\mu\nu} + {\rm Tr} (\nabla^\mu \phi)^\dagger 
 \nabla_\mu \phi  + \ldots 
\Bigr) 
\fqr 
has a $N$-dependence in the correlator of gauge invariant composite 
operators $O_{\{i\}}({\vec x}_i)$ in perturbation theory at small 't 
Hooft coupling $\lambda=g^2N$ of 
\bqr  
\langle \prod_{i=1}^n O_{\{i\}}({\vec x}_i) \rangle = N^{2(1-g)}  
  \lambda^{p} C_{\{i_j\}} ({\vec x}_i) \ .
\label{lambdaexpand}
\fqr 
The chiral primary operators (composed of a symmetric traceless representation 
of $n$ chiral fields $\phi^{i}$) relevant to \rf{lambdaexpand} are 
normalized so that 
\bqr  
O^{\{i_j\}}({\vec x}) = N \lambda^{-{n\over 2}}~ {\rm Tr} \Bigl[
   \phi^{\{i_1}({\vec x}) \ldots \phi^{i_n\}}({\vec x}) \Bigr] \ . 
\fqr 
The dependence in \rf{duallimit} arises as a reexpansion of the $N$ and 
$\lambda$ series, with $N$ large and $\lambda/N^2$ fixed, of the 
microscopic theory with graphs of weight \rf{lambdaexpand}.  Although 
the weak coupling limit $\lambda/N^2<< 1$ does not permit a description 
in terms of a microscopic $\N=4$ gauge theory because there are no 
Feynman diagrams with the appropriate powers of $\lambda/N^2$, it does 
have a perturbative description in terms of holographic IIB superstring 
theory on $AdS_5\times S^5$.   More specifically, it is described by the 
genus $g$ calculations obtained directly from the $\N=4$ topological 
string, which have also been used to compute similar terms in 
the IIB superstring \cite{Berkovits:1995vy} with conjectured truncated 
complete results.  

The correlation function, as an expansion in $N$, from the weak 
coupling side has the form based on the perturbative structure 
of Feynman diagrams, 
\bqr  
\langle \prod_{i=1}^4 O_{\Delta_i}({\vec x}_i) \rangle = N^2 F_0(\lambda) 
+ F_1(\lambda) + {1\over N^2} F_2(\lambda) + \ldots 
\label{pertexpansion}
\fqr 
Simply exchanging $\lambda\rightarrow {N^2\over\lambda}$ to obtain 
a dual limit, 
\bqr  
\langle \prod_{i=1}^4 O_{\Delta_i}({\vec x}_i) \rangle = N^2 F_0(N^2/\lambda) 
+ F_1(N^2/\lambda) + {1\over N^2} F_2(N^2/\lambda) + \ldots \ , 
\label{dualexpression}
\fqr 
via S-duality of $\N=4$ super Yang-Mills theory gives the same function 
as in \rf{pertexpansion}.  This exchange of $\lambda$ with its dual 
would indicate that the dual 't Hooft limit has a power series expansion 
generically, if the original expansion does in both limits $\lambda 
\rightarrow 0$ and $\lambda\rightarrow\infty$ based on perturbation 
theory.   However, the strongly coupled limit of 
$F_j(\lambda)$ is subject to have a limit, i.e. at infinite $\lambda$ and 
finite $N$ the terms in the  series have a maximum power of $\lambda$.  For 
example, the four-point function from the string point of view generate 
at infinite $N$ the structure of order 1 with a sub-leading contribution 
of order $\lambda^{-3/2}$.  The existence of the strongly coupled limit 
in the large $N$ limit requires additional input then just the functional 
form in \rf{pertexpansion} together with its dual relation \rf{dualexpression}.  

The analagous statement made for the dual limit requires the re-expansion 
of the expression in \rf{dualexpression} at large values of $\lambda/N^2$ 
together with the incorporation of instanton effects.  The first instanton, 
for example, has the form 
\bqr  
{1\over N^{1/2}} e^{-N/\lambda} \rightarrow 
 {1\over N^{1/2}} e^{-N(\lambda/N^2)} \ ,  \label{inst}
\fqr 
and does not fit into the expansion form of \rf{pertexpansion}, which 
is based on the topology of Feynman diagrams.  The consistency of the 
dual 't Hooft limit requires that the instanton expansion is compatible 
with the expansion in $\lambda/N^2$.  We require that in the large  
$N$ limit that the power series does not generate an ever increasing 
number of terms with factors $N^{2p}$.  A consistency check is to 
show that potential of this form terms could not exponentiate into an instanton 
type of effect in the dual limit (at sub-leading order in $N$).  

The genus selection of the $\Box^k R^4$ terms in the low-energy 
expansion in the limit ${\tilde\lambda}=\lambda/N^2$ fixed 
is related by supersymmetry to the amplitudes $R^4 H^{4g-4}$ in the  
string in \cite{Berkovits:1998ex,Berkovits:1995vy} because the same 
order in the genus is naturally obtained; a holographic formulation of 
this $\N=4$ superstring generates the coupling constant dependence of the  
dual field theory description.  By virtue of $S$-duality, 
the existence of the series with 
\rf{duallimit} 
in $\N=4$ super Yang-Mills theory
at finite values of $\lambda/N^2$ translates into the 
perturbative truncation property of the genus expansion.   

\section{Dual Planar Limit and Genus Truncation} 

\setcounter{equation}{0}

The structure of the series in \rf{fseries} is such that the $\N=4$ super 
Yang-Mills theory planar expansion at finite $\lambda$ exists under the 
conditions outlined in \rf{dualplanar}.  As we take $N$ large in the 
$\lambda/N$ fixed limit of (\ref{weak}), 
the leading order supergravity approximation is 
obtained because the individual terms in \rf{fseries} go to zero as 
$1/N\rightarrow 0$.  Similarly, at large $\lambda$ and large $N$, the 
leading term of every $1/N^{2p}$ contribution to
(\ref{leadingduallimit}) is given by
\bqr  
\langle \prod_{j=1}^4  T^{a_j b_j}({\vec x}_j) \rangle_p 
 = N^{-2p}~ {\lambda^{p-{1\over 2}}} a_p^{(2p-2)}({\vec x}_i) + 
 {\cal O}(\lambda^{p-{3\over 2}}) \ ,  
\label{nonrenorm} 
\fqr 
with similar structure in the generic correlation functions.  For a 
given factor $N^{2p}$ the form indicates a maximum power $\lambda^{p-{1/2}}$ 
in the $\lambda$ series.  The leading $\lambda$ power indicates, in effect, 
an infinite number of non-renormalizations for
the dependence on the sub-leading $1/N^p$ terms
at strong coupling (similarly to the perturbative series at 
weak coupling in an expansion in $\lambda$).  

In this section we will investigate the possibility of higher genus 
occurences above $g_{\rm max}$ in string theory; this would lead to an infinite number 
of higher order terms in \rf{nonrenorm} which are of higher order
in $\lambda$. 
The duality between IIB superstring theory and $\N=4$ gauge theory has 
in the past given a non-perturbative insight into the latter. In
this analysis we shall extract information about the perturbative string 
from the gauge theory through the dual description. Specifically, 
by assuming that the existence of the expansion in $\lambda_D$ in the S-dual of the 
't Hooft limit may be proved within field theory, we show that contributions
to the string perturbation theory expansion for which $g> g_{\rm max}$ 
must be absent.  In the dual planar 
expansion, $\lambda/N^2={\rm fixed}$ and $N\rightarrow\infty$, the generic
contribution from higher genus terms could generate terms that diverge 
individually in $N$ of increasing power at a fixed power of $\lambda/N^2$.  
For example, power counting at genus two relevant to the $R^4$ term gives
\bqr  
{\lambda\over\lambda^{1/2}} \bigl({\lambda\over N^2}\bigr)^2 = 
N \bigl({\lambda\over N^2}\bigr)^{5/2} \ , 
\fqr 
that at genus four relevant to $\Box^2 R^4$,   
\bqr  
{\lambda^2\over \lambda^{1/2}} \bigl({\lambda\over N^2}\bigr)^4 = N^3
\bigl({\lambda\over N^2}\bigr)^{11/2} \ , 
\fqr 
and as another example, the genus three contribution to the $R^4$ term, 
\bqr  
{\lambda^2\over\lambda^{1/2}} \bigl({\lambda\over N^2}\bigr)^3 = 
N^3 \bigl({\lambda\over N^2}\bigr)^{9/2} \ . 
\fqr 
The general counting at order $2p$ in derivatives and at genus $g$ 
follows from that in
\rf{fseries}, i.e. 
\bqr  
\lambda^{-3/2-p/2} \bigl({\lambda\over N}\bigr)^{2g} = N^{2g-3-p} 
\bigl({\lambda\over N^2}\bigr)^{2g-3/2-p/2} \ .
\label{maxNcount}
\fqr 
For $g\geq {1/2}(p+2)$ the factors of $N$ increase relative to the 
leading $N$ term at fixed $\lambda/N^2$.  (Note that there is an 
overall factor of $N^2$ which has not been included in the normalization 
of the counting in \rf{maxNcount}). 

\begin{table}  
\caption{
Non-analytic dependence tied to the leading even genus contributions.  The 
former contributes at higher order in $\alpha'$ at the same genus order.} 
\begin{center}
\begin{tabular}{|l|l|l|l|l|l|r|} \hline 
{\em } &  $g=0$ & $g=1$ & $g=2$ & $g=3$ & $g=4$ & $\ldots$ \\ \hline 
$R^4$        & & $\surd^\star$ & & & & \\ \hline  
$\Box R^4$   & & $\surd_0$  & & & & \\ \hline  
$\Box^2 R^4$ & & $\ldots$ & $\surd^\star$ & & &  \\ \hline 
$\Box^3 R^4$ & & & $\surd_2$ & & & \\ \hline 
$\Box^4 R^4$ & & & $\ldots$  & $\surd^\star$ & & \\ \hline 
$\Box^5 R^4$ & &  &  & $\surd_4$ & & \\ \hline 
$\Box^6 R^4$ & &  &  & $\ldots$ & $\surd^\star$ & \\ \hline
$\cdots$ & & & & & & \\ \hline     
\end{tabular} 
\end{center} 
\end{table} 

Contributions to $\Box^k R^4$ in the genus expansion with $g>g_{\rm max}$ 
generate further $\lambda/N^2$ powers to \rf{leadingduallimit} but 
with additional powers of $N$ (or $\lambda$ for fixed $\lambda_D=N^2 
/\lambda$).  We label these terms by ${\tilde f}({\vec x}_i)$, with 
explicit form   
\bqr 
{\tilde f}({\vec x}_i) &=& \sum_{p=1}^\infty \sum_{m_p=1}^{2p-2} 
 \lambda^{-{1\over 2}+{m_p\over 2}} \Bigl({\lambda\over N^2}\Bigr)^p 
 a_p^{(2p-2-m_p)}({\vec x}_i)  \ .
\fqr 
The generalization to all of the even $k$ terms in \rf{fseries} is   
\bqr 
{\tilde f}({\vec x}_i) = \sum_{a=0}^\infty \sum_{p=1}^\infty \sum_{m_p=1}^{2p-2} 
 \lambda^{-{1\over 2}} N^{m_p-2a} \Bigl({\lambda\over N^2}\Bigr)^{p+ 
 {m_p\over 2}-a} a_p^{(2p-2-m_p+2a)}({\vec x}_i)  \, .
\label{highergenus} 
\fqr  
The sum over $m_p$ represents the possible higher genus contributions 
beyond those of $g_{\rm max}$, and the sum over $a$ denotes the sub-leading 
in $\lambda$ terms in \rf{fseries}.   In \rf{highergenus} either each 
term diverges as $N$ becomes large or is suppressed in $N$.  Therefore, 
consistency with the convergence of the series in (\ref{leadingduallimit}) 
in the S-dual of the planar limit, which we assume to exist in the field 
theory, requires that the divergent terms in (\ref{highergenus}) are absent 
and implies that there are no terms with $g> g_{\rm max}$ in the string 
theory expansion. Moreover such terms would not be of the form (\ref{inst}) 
required by compatibility with the instanton expansion.

\section{Tree-Level: Infinite $N$} 

\setcounter{equation}{0}

The explicit summation of all of the leading derivative structures 
in \rf{sform} contributing at infinite $N$ may be extracted from 
the tree-level superstring scattering amplitude.  At tree-level in 
the graviton scattering amplitude there are no space-time fermions 
contributing to the amplitude, yet the boundary dual correlator is 
that of a supersymmetric field theory at finite coupling constant; 
the supersymmetry of the infinite $N$ result is encoded in the isometry 
structure of the bulk spacetime related to the $R$-symmetry of the gauge 
theory.  We shall compare the result with the expected unitarity and 
$\lambda$ structure of the $\N=4$ super Yang-Mills theory.  The general 
covariantized $n$-point function of gravitons in IIB string theory at 
tree-level gives rise to a Koba-Nielsen like form of the correlation 
function $\langle \prod_{j=1}^n T_{a_j b_j}({\vec x}_j) 
\rangle$ at infinite $\N$  and finite $\lambda$.  The integral result 
of the latter is expressed as a one-parameter integration (over the 
holographic dimension $z_0$) which contains the physical features of 
the correlator in $\N=4$ super Yang-Mills theory as described in this 
section.  Similar amplitudes arising from \rf{fiveform} and \rf{multipoint} 
would need to be computed in order to obtain the complete large $N$ 
result in the background. 

The four-point graviton IIB scattering amplitude (in flat space) is 
described by the well-known amplitude in string frame,  
\bqr  
A_4 = e^{-2\phi} {R^4\over stu} {\Gamma(1-\alpha's) \Gamma(1-\alpha't) 
 \Gamma(1-\alpha'u)\over \Gamma(1+\alpha's) \Gamma(1+\alpha't) 
\Gamma(1+\alpha'u)} \ ,  
\label{fourpointtree}
\fqr 
and alternatively as 
\bqr  
A_4 = e^{-2\phi} {R^4\over stu} \exp{\sum_{k=1}^\infty {2\zeta(2k+1)\over 
2k+1} (\alpha')^{2k+1} \bigl( s^{2k+1}+t^{2k+1}+u^{2k+1} \bigr)} \ , 
\fqr 
which may be found by summing over the four-punctured tori of the 
correlation of four graviton vertex operators.  It is interesting 
to note that upon $\alpha'\rightarrow -\alpha'$ (corresponding to 
$\lambda^{1/2}\rightarrow -\lambda^{1/2}$) the ratio of gamma functions 
in \rf{fourpointtree} is inverted.  

The general $n$-point amplitude of gravitons with individual polarizations 
$\epsilon_{\mu\bar\mu}$ decomposed into $\epsilon_\mu$ and ${\bar\epsilon}_\mu$ 
at tree-level is 
\bqr  
A_n(k_i,\epsilon_i) = e^{-2\phi} \int d\mu_n^s ~\prod_{i<j}  
 \vert e^{-\alpha' k_i\cdot k_j G_{ij} + \epsilon_i \cdot k_j D^+_i 
G_{ij} +(i\leftrightarrow j)+ \epsilon_i \cdot\epsilon_j D^+_i D^+_j G_{ij} + 
 \epsilon_i \cdot{\bar\epsilon}_j D^+_i D^-_j G_{ij}}\vert ~  
 \vert_{\rm multi} \ ,   
\label{KNpoint}
\fqr 
found directly by the known path integration over the $n$-punctured 
super sphere; the form in \rf{KNpoint} is to be expanded multi-linearly 
in the polarizations. The holomorphic half of the world-sheet scalar superfield 
Greens function is 
\bqr  
G_{ij} = - \ln(z_i - z_j + \theta_i\theta_j) \ , 
\fqr 
and in \rf{KNpoint} we have the measure 
\bqr 
d\mu_n^s = \vert z_i - z_j \vert^2 \vert z_j - z_k\vert^2 
 \vert z_k - z_l\vert^2 ~ d\rho_n(\theta) \prod_{m=1,\neq i,j,k}^n d^2 z_j  
\ ,
\fqr 
over the integration of the vertex insertion points with three 
fixed points (and the measure over the fermionic components $\theta$). 

The expansion in $\alpha'$ of \rf{fourpointtree} generates the 
low-energy expansion of the covariantized S-matrix, 
\bqr  
S= e^{-2\phi} \int d^{10}x \sqrt{g} \Bigl[ {1\over \alpha'^3 stu} R^4 
+ \sum_{k=1}^\infty g_k (\alpha'\Box)^k R^4 \Bigr] \ , 
\label{derexpansion} 
\fqr 
which is the first in an infinite series in the genus expansion.  
The derivatives and their order in \rf{derexpansion} are arranged 
in accord with the expansion of the four-point function in 
\rf{fourpointtree}.  

\begin{figure}
\begin{center}
\epsfig{file=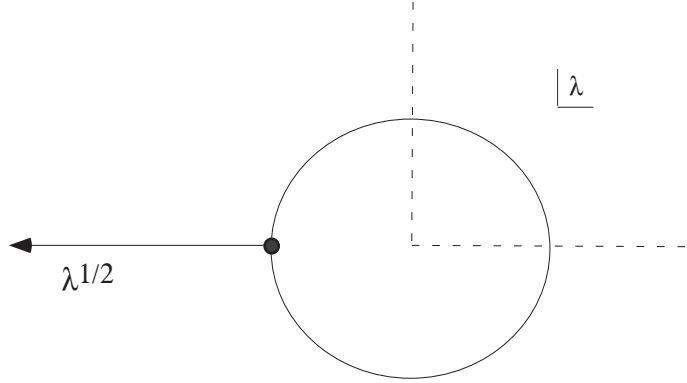,height=5cm,width=9cm} 
\end{center} 
\caption{The complex $\lambda$-plane.  The expansion of the four-point 
correlator around $\lambda=\infty$ is double-valued and indicates a branch 
cut.  We represent it here from $\lambda=-1$ to the point at (minus) infinity.} 
\end{figure} 

The explicit background expansion of the covariantized four-graviton 
scattering amplitude resummed in derivatives produces the Gamma 
function form in \rf{fourpointtree}.  Indeed, the ratio of Gamma 
functions is of the form, $\alpha' s_{\rm bdy}\rightarrow (\alpha'/R_{AdS}^2) 
~z_0^{2} ~(k_1+k_2)\cdot (k_1+k_2)$, 
\bqr 
{\Gamma(1-\lambda^{-1/2} s) \Gamma(1-\lambda^{-1/2} t) 
 \Gamma(1-\lambda^{-1/2} u) \over \Gamma(1+\lambda^{-1/2} s)
 \Gamma(1+\lambda^{-1/2} t) \Gamma(1+\lambda^{-1/2} u) } \ ,
\fqr 
where the invariants represent the differential operators in the 
$(1,2)$, $(1,3)$ and $(2,3)$ channels such that 
\bqr  
s= z_0^2 \bigl( 
{\partial\over\partial z_0} {\partial\over\partial z_0} + 
 ({\partial\over\partial {\vec x}_1}+{\partial\over\partial{\vec x}_2}) \cdot
({\partial\over\partial{\vec x}_1}+{\partial\over\partial{\vec x}_2}) 
 \bigr) \ ,  
\label{differential}    
\fqr 
where the Fourier expansion would generate $(k_1+k_2)^2$ in the boundary 
components from the partial derivatives.    
The explicit evaluation of the generating function, or one-parameter 
integral, for the $N_c\rightarrow\infty$ and finite $\lambda$ correlation 
function requires the Gamma function with the differential argument in 
\rf{differential} 
\bqr 
\mathop{\rm lim}_{N\rightarrow\infty} \langle 
\prod_{j=1}^4 T^{a_j b_j}({\vec x}_j) \rangle = N^2  
 \int {dz_0 d^d{\vec z}\over z_0^{d+1}}  ~ {1\over s t u}
 {\Gamma(1-\lambda^{-1/2} s) \Gamma(1-\lambda^{-1/2} t) 
 \Gamma(1-\lambda^{-1/2} u) \over \Gamma(1+\lambda^{-1/2} s)
 \Gamma(1+\lambda^{-1/2} t) \Gamma(1+\lambda^{-1/2} u) } 
\fqr 
\bqr 
 \times \Bigl[ \prod_{i=1}^4 G^{\{a_ib_i\}}{}_{\{\mu_i\nu_i\}} 
 ({\vec x}_i; z_0,{\vec z}) \Bigr] \prod_{j=1}^4 \bigl\{ {\delta\over
\delta {\hat g}_{\mu_j\nu_j} (z_0,{\vec z})} \bigr\} \Bigl[ R^4 \Bigr]  
\vert_{g_{AdS_5\times S^5}}~ 
 \ ,  
\label{infiniteN}
\fqr 
together with the propagation associated withe the graviton intermediate 
lines; the expression in \rf{infiniteN} can be understood as an infinite 
expansion given in \rf{derexpansion} of the $k\geq 0$ terms, but resummed. 
We have included only the four-point graviton scattering in \rf{infiniteN}, 
and there are further contributions from the background effects associated 
with the terms in \rf{fiveform} and \rf{multipoint}.    
The integration over the ${\vec z}$ components in \rf{infiniteN} is 
straightforward after Fourier transforming to momentum space.  The $\lambda$
dependence is manifest in \rf{infiniteN}.  We comment on three points 
regarding the integral form in \rf{infiniteN}.  

First, because the Gamma function structure in the integrand is inverted 
under $\lambda^{-1/2} \rightarrow  -\lambda^{-1/2}$ the one-parameter 
integral (in $z_0$) representing the infinite $N_c$ correlator in 
\rf{infiniteN} satisfies the reflection property, 
\bqr  
\mathop{\rm lim}_{N\rightarrow\infty} \langle 
\prod_{j=1}^4 T_{a_j b_j}({\vec x}_j) \rangle~
\vert_{e^{i\theta}\vert\lambda\vert^{-1/2}} = \Bigl[ \mathop{\rm
lim}_{N\rightarrow\infty} \langle 
\prod_{j=1}^4 T_{a_j b_j}({\vec x}_j) \rangle~
\vert_{e^{-i\theta}\vert\lambda\vert^{-1/2}} \Bigr]^\star
\label{cutform} \ ,
\fqr 
for $\theta<\pi$.  The inclusion of the massless graviton exchange is 
one term in the series and by normalization is $\lambda$ independent 
in the field theory; it does not change the analysis.  As a function 
of the complex parameter $\lambda^{-1}=e^{i\phi} \vert\lambda\vert^{-1}$, 
it has a branch cut in  the complex $\lambda$-plane; taking $\lambda^{-1} 
=e^{i\phi} \vert\lambda\vert^{-1}$ and sending $\phi\rightarrow\phi+2\pi$ 
the Gamma functions in \rf{infiniteN} get inverted.  After rotating twice 
in the complex plane the correlation function is invariant, 
\bqr  
\mathop{\rm lim}_{N\rightarrow\infty} \langle \prod_{j=1}^4 
T_{a_jb_j}({\vec x}_j) \rangle~ \vert_{e^{4\pi i} \lambda}  
= \mathop{\rm lim}_{N\rightarrow\infty} \langle \prod_{j=1}^4 
T_{a_jb_j}({\vec x}_j) \rangle~ \vert_{\lambda} \ .
\label{monodromy}
\fqr 
The square root of the coupling is what is required to make the dual 
correlator form in \rf{infiniteN} invariant after two rotations in the 
complex plane.  In perturbation theory the correlator is termwise a 
function of $\lambda$ and so this monodromy structure does not appear 
in the perturbative series without resumming.  At small $\lambda$ 
coupling in perturbation theory there is no phase associated with the 
termwise rotation under $\lambda\rightarrow e^{2\pi i}\lambda$.  A 
square root branch cut reflecting the double cover of the $\lambda$ 
plane possibly exists from $\lambda=-1$ to the point at infinity, 
illustrated in figure 5.  There is a $Z_2$ ambiguity in relating the dual 
strongly coupled limit of the $\N=4$ gauge theory with that of the IIB 
superstring because of the branch cut in this description.  The relation 
\rf{vardefin} together with the non-invariance of string theory as 
$\alpha'\rightarrow -\alpha$ is the origin of the ambiguity.  

Second, the free-field limit of the correlation function 
in ${\cal N}=4$ super Yang-Mills theory is related to the Gross-Mende 
limit of the string scattering.  The evaluation may be approximated 
by the application of Stirling's formula, 
\bqr 
\Gamma(z)=\sqrt{2\pi} z^{z-{1/2}} e^{-z} ~ 
 \bigl[1 + {\cal O}({1\over z})\bigr] \ , 
\fqr   
which might enable a comparison with the free-field theory limit 
obtained at $\lambda=0$.

\begin{figure}
\begin{center}
\epsfig{file=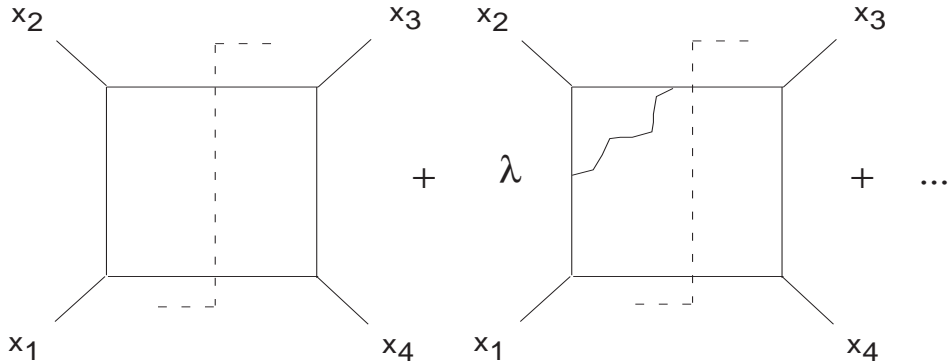,height=5cm,width=13cm} 
\end{center} 
\caption{The unitarity properties of the correlator in the two-particle 
channel at finite $\lambda$ has a non-trivial structure found by 
summing and reexpanding the perturbative series.  The string tree 
exchange in \rf{infiniteN} generates unitarity cuts after integration 
over $z_0$, found alternatively by resumming the low-energy scattering. } 
\end{figure} 

Our third point refers to the unitarity structure and its comparison 
with the dual field theory result. 
As in \rf{effvertex} the holographic integral representation is found 
by four variations of the Einstein frame result in \rf{fourpointtree} 
together with an integration over the bulk points.  The massless 
graviton contribution is obtained at low-energy by integrating over 
the supergravity modes directly in the AdS background.  In the approximated 
form given in \rf{derexpansion} the unitarity cut in the two-particle 
$(12)$ channel (analyzed in detail in \cite{Chalmers:1999wu}) found by 
\bqr  
\mathop{\rm Im}_{s<0} ~\langle \prod_{j=1}^4  
 T_{a_j b_j}({\vec k}_j) \rangle  \ , 
\fqr  
in the regime, 
\bqr  
({\vec k}_1+{\vec k}_2)^2<0 \qquad ({\vec k}_1+{\vec k}_4)\geq 0, \quad 
 ({\vec k}_2+{\vec k}_4)^2 \geq 0 \ , 
\fqr 
is not clear because the four-point graphs do not individually in perturbation 
theory possess the poles to extract the imaginary part.  However, after 
resumming the expansion, i.e. in the form of \rf{fourpointtree}, the 
poles at arbitrary mass level (and intermediate $\lambda$ are obtained).  
Indeed, individually from the weak-coupling series in $\N=4$ super Yang-Mills 
theory there are unitarity cuts in the two-particle channel generically 
at every order in $\lambda$, illustrated in figure 6.  They are found 
by integrating over $z_0$ and summing the series in $\alpha'$; the poles 
of the gamma function in \rf{fourpointtree} are obtained by integrating 
over $z_0$ at fixed momenta.  The integration over $z_0$ in \rf{infiniteN} 
in principle extracts the unitarity cuts of the complete $\N=4$ super 
Yang-Mills result (expanded around $\lambda\rightarrow\infty$ as opposed to 
$\lambda\rightarrow 0$) \cite{Chalmers:1999xj}. 

\setcounter{equation}{0}

\section{Space-Time Structure} 
\label{OPEsec}

Here we calculate the explicit dependence of the contact diagram contributions 
to the stress tensor four-point function given by (\ref{genexpand}). 
This leads to expressions involving Feynman parameter integrals which give 
rise to power series in the cross-ratios in general.  For analysing the OPE 
structure of these expressions we evaluate these  integrals in the limit when 
two of the four points approach each other. In Section \ref{OPEsec}.1 we discuss 
the OPE of two stress tensors for a general conformal field theory in $d$ 
dimensions. In Section \ref{OPEsec}.2  we turn to the OPE behaviour of the 
four-point function arising from (\ref{genexpand}).  In Sections \ref{OPEsec}.3 
and \ref{OPEsec}.4 we analyze the triangle integrals arising in the discussion 
of the space-time structure for arbitrary separation of the four points. 
Related OPE investigations may be found in \cite{Liu:1999th,Anselmi1,Dhoker1,Petkou1}.

\subsection{Operator Product Expansion}

For a general $d$-dimensional CFT, the stress tensor OPE contains terms of 
the form $(\vec{x}_{21} \equiv \vec{x}_1 - \vec{x}_2)$
\begin{eqnarray} 
\lefteqn{T_{a_1 b_1}(\vec{x}_1) T_{a_2 b_2}(\vec{x}_2)} \,  && \nonumber\\
& \sim&  \, 
t^{O}_{a_1 b_1, a_2 b_2}(\vec{x}_{21}) \, O_0(\vec{x}_2)
+ t^{V}_{a_1 b_1, a_2 b_2, k}(\vec{x}_{21}) V_k(\vec{x}_2 )
+ t^{T}_{a_1 b_1, a_2 b_2, kl}(\vec{x}_{21}) 
\, T_{kl} (\vec{x}_2) \, + \cdots \, , \label{OPE}
\end{eqnarray} 
where $O_0$ is a dimensionless scalar and $V_k$ is a conserved
vector current of dimension $d-1$.  The terms shown explicitly are all 
leading terms in the sense that the tensors $t$ are of dimension $t \sim
|x_{21}|{}^{\lambda}$ which $ \lambda \geq d$, 
such that they may potentially give rise
to poles by virtue of \cite{Gelfand}
\begin{equation}
\frac{1}{|\vec{x}_{12}|^{\lambda}} \sim \frac{1}{d + 2n -  \lambda}
\frac{1}{2^{2n} n! } \frac{2 \pi^{d/2}}{\Gamma(d/2+n)} \,  (\partial^2)^n
\delta^d (\vec{x}_{12}) \, ,  \label{limit2}
\end{equation}
where $n=0,1,2 \dots$. 
There are leading contributions to (\ref{OPE}), where `leading' is 
used in the sense discussed above, for example scalar operators of dimension
$\eta \leq d$ or from a two-form operator $F_{ab}$ of scale dimension $d-2$.
For our discussion here however it is sufficient to consider the terms shown.
Of course there are also non-leading terms contributing to (\ref{OPE}). 
As discussed extensively in \cite{Osborn:1994cr},
\cite{Erdmenger:1997yc} , the tensors $t$ play a crucial role
in constructing conformal three-point functions. In
\cite{Osborn:1994cr}
it was shown
that for a $d$-dimensional CFT, the three-point function involving
two stress tensors and a third operator $\O_J$ of arbitrary spin $J$
and dimension $\eta$ (here $\O_J \in \{ O_0, V_k, T_{kl} \}$) 
is given to leading order by  
\begin{eqnarray}
\langle T_{a_1 b_1}(\vec{x}_1) T_{a_2 b_2}(\vec{x}_2)
T_{a_3 b_3}(\vec{x}_3) \rangle &=& \frac{ \I^T{}_{a_2 b_2, kl}
(\vec{x}_{21}) \I^T{}_{a_3 b_3, mn}
(\vec{x}_{31}) }{|\vec{x}_{21} |^{2d} 
|\vec{x}_{31} |^{2d}} t^J{}_{kl,mn, J} (X) \, , \label{TTI}
\end{eqnarray} 
where 
\begin{eqnarray} 
X_k \equiv \frac{{\vec{x}}_{21}{}_k}{|\vec{x}_{21} |^2}
- \frac{\vec{x}_{31}{}_k }{|\vec{x}_{31} |^2} \quad , \,
\I^T{}_{a b, kl}(x) \equiv \E_{ab,mn} 
 I_{mk}(x) I_{nl}(x) \, . \label{it}
\end{eqnarray}
Here
\bqr  \label{inv}
I_{ab}(\vec{y}) = \delta_{ab} - 2 {\vec{y}_a \vec{y}_b\over |\vec{y}|^2}  
\fqr 
is the defining representation of the inversion and
\begin{equation}
\E_{kl,mn} = \half ( \delta_{km}\delta_{ln}
+ \delta_{lm}\delta_{kn}) - {\ts \frac{1}{d}} \delta_{km}\delta_{ln}
\end{equation}
the projector onto traceless symmetric tensors which satisfies 
\bqr 
\E_{ij,kl} I_{lm} = \delta_{ik} I_{jm} + \delta_{jk} I_{im} 
 -{1\over d} \delta_{ij} I_{km} \ .  
\label{PIidentity} 
\fqr 
It was shown in \cite{Osborn:1994cr} that the form (\ref{TTI})
for the conformal three-point function is unique in the sense that 
it is both necessary and sufficient. The tensor $t^J$ has to satisfy 
constraints arising from symmetry and stress-tensor conservation. 

Let us now consider the stress tensor four-point function.  When
$\vec{x}_1 \rightarrow \vec{x}_2$ we have ($\vec{x}_{21}\equiv
\vec{x}_2 - \vec{x}_1 $)
\begin{eqnarray}
\lefteqn{
\langle T_{a_1 b_1}({\vec x}_1) T_{a_2 b_2}({\vec x}_2) 
 T_{a_3b_3}({\vec x}_3) T_{a_4b_4}({\vec x}_4) \rangle } \hspace{1cm} \nonumber\\
&=& 
t^O_{a_1 b_1, a_2 b_2}(\vec{x}_{21})
\langle O_0({\vec x}_2) 
 T_{a_3b_3}({\vec x}_3) T_{a_4b_4}({\vec x}_4) \rangle \,  
\\ && + 
t^V_{a_1 b_1, a_2 b_2, k}(\vec{x}_{21})
\langle V_{k}({\vec x}_2) 
 T_{a_3b_3}({\vec x}_3) T_{a_4b_4}({\vec x}_4) \rangle \nonumber\\
&& + \,
t^T_{a_1 b_1, a_2 b_2, kl}(\vec{x}_{21})
\langle T_{kl}({\vec x}_2) 
 T_{a_3b_3}({\vec x}_3) T_{a_4b_4}({\vec x}_4) \rangle
+ \cdots \, , \nonumber\label{4ptOPE}
\end{eqnarray}
where the three point functions are of the form (\ref{TTI}). 
As was shown in \cite{Osborn:1994cr}, the three point function involving the
dimensionless scalar $O_0(\vec{x}_2)$ is independent of $\vec{x}_2$ and is
proportional to the two-point function
\begin{equation}
\langle O_0({\vec x}_2) 
 T_{a_3b_3}({\vec x}_3) T_{a_4b_4}({\vec x}_4) \rangle 
\propto \langle T_{a_3b_3}({\vec x}_3) T_{a_4b_4}({\vec x}_4) \rangle 
= C_T \, \frac{ \I^T_{a_3b_3,a_4b_4}(x_{12})}{ (x_{12}{}^2)^d} \, .
\end{equation}
Similarly $t^O_{a_1b_1 a_2 b_2}(\vec{x}_{12})$ 
is also proportional to the two-point function,
\begin{equation}
t^O_{a_1 b_1 a_2 b_2}(\vec{x}_{12}) \propto 
\langle T_{a_3 1 b_1}({\vec x}_1) T_{a_2 b_2}({\vec x}_2) \rangle \, ,
\end{equation}
such that the dimensionless scalar in the OPE leads to the contribution
to the four-point function which factors into two two-point functions,
\begin{equation}
\langle T_{a_1 b_1}({\vec x}_1) T_{a_2 b_2}({\vec x}_2) 
T_{a_3b_3}({\vec x}_3) T_{a_4b_4}({\vec x}_4) \rangle_{O}
\propto 
\langle T_{a_1 b_1}({\vec x}_1) T_{a_2 b_2}({\vec x}_2) \rangle \, \langle 
 T_{a_3b_3}({\vec x}_3) T_{a_4b_4}({\vec x}_4) \rangle \, .
\end{equation}
As discussed in \cite{Erdmenger:1997yc}, $\langle T_{a_1 b_1}({\vec x}_1) 
T_{a_2 b_2}({\vec x}_2) \rangle$ contains a pole due to (\ref{limit2}).  As 
far as $t^T_{a_1 b_1 a_2 b_2 kl}$ in (\ref{TTI}) is concerned, it was shown in 
\cite{Erdmenger:1997yc} that although the dimension is $t^T_{a_1 b_1 a_2 b_2 kl} 
\sim |x_{21}|{}^{-d}$, such that a pole of type (\ref{limit2}) might be expected, 
the tensorial structure ensures that there is no such pole if energy-momentum 
conservation is imposed. We expect a similar result to hold for 
$t^V_{a_1 b_1 a_2 b_2 k }$.

\subsection{String theory contributions to the stress tensor 
four-point function}

We are going to show that the contribution to the stress tensor four-point
function (\ref{genexpand}), which corresponds to contact diagrams for the graviton scattering,
does not contain any leading OPE term in the sense
that when $x_1 \rightarrow x_2$, there is no term involving $|\vec{x}_{21}|{}^{-\lambda}$
with $\lambda \geq d$,  which implies that the tensors $t^{J}$ discussed 
above are identically zero for this contribution. For the explicit
evaluation of (\ref{genexpand}) we generalize techniques developed in
\cite{Frolov}. At first we consider the case $k=0$ in (\ref{genexpand}),
which corresponds to the well-known $R^4$ term, and discuss the extension
to $k \neq 0$ and to the five-form terms later on. 

For the $AdS_5$ geometry we use the conventions of \cite{Freedman:1999tz}.
We use the $AdS_5$ bulk-to-boundary propagator
in the the Donder gauge \cite{Tseytlin}, which is given by
\bqr 
G_{\mu\nu,}{}_{ab}({\vec x};z_0,{\vec z}) = \Bigl[ (d+1) 
{\Gamma(d-1)\over \pi^{d/2} \Gamma({d\over 2})}\Bigr]    
 {z_0{}^{d-2}\over \left(z_0^2 + \vert {\vec z}-{\vec x}\vert^2\right)^d} 
 I_{\mu k} (z-{\vec x})  I_{\nu l} (z-{\vec x}) \E_{kl}{}_{ab} \ . 
\label{bulkboundary}  
\fqr 
Our conventions for the connection and for the covariant derivatives are
\bqr 
\Gamma_{\mu\nu}{}^\kappa = -{1\over z_0} \Bigl[ \eta_\mu^0 \eta^\kappa_\nu 
+ \eta_\nu^0 \eta^\kappa_\mu - \eta^{0\kappa}\eta_{\mu\nu} \Bigr] \ ,
\fqr  
\bqr  
\nabla_\mu V_\nu = \partial_\mu V_\nu - \Gamma_{\mu\nu}{}^\kappa V_\kappa \ , 
\fqr 
where for the AdS metric we have $g_{\mu\nu}={1\over z_0^2}\eta_{\mu\nu}$.
For evaluating the vertex given by (\ref{genexpand}) we make use of the fact that 
for the metric fluctuations around the background we have
\begin{equation}
\delta R_{\mu \sigma \rho \nu} = \nabla_{[\mu}\nabla_{[\nu}\delta g_{\rho ] \sigma ] }\,
 ,
\end{equation}
where the antisymmetrization is in pairs $[\mu ,\sigma]$ and $[\nu, \rho]$. 
We find that 
\begin{eqnarray} 
\nabla_{[\mu}^z \nabla_{[\nu }^z  
 G_{\rho]\sigma];a b}(z,{\vec x}) &=& \Bigl[  
{\Gamma(d+2)\over \pi^{d/2} \Gamma({d\over 2})}\Bigr] \, 
{z_0^{d-4}\over (z-{\vec x})^{2d}} \nonumber\\
&& \hspace{-2cm}
\times \Bigl[ I_{0[\mu} I_{\sigma]k} 
 I_{0[{\nu}} I_{{\rho}]l} \E_{kl, a b}
- { \frac{1}{d}} \eta_{[\mu [\nu} I_{\sigma]k} I_{\rho]l} 
\E_{kl,ab} \Bigr] (z-{\vec x}) \, , 
\label{covidentity}
\end{eqnarray}
the form of which is of particular importance for the evaluation of the four-point
function. This result is very similar to a result of \cite{Freedman:1999tz} 
for the bulk-to-boundary gauge field propagator $G^G$ for which 
\begin{eqnarray}
G^G_{\mu i}(z,\vec{x}) &=& C^d \frac{z_0^{d-2}}{(z - \vec{x})^{2 (d -
1)}}\,I_{\mu i}(z - \vec{x})\, ,\nonumber\\
\partial_{[\mu}G^G_{\nu ]i}(z,\vec{x}) &=& (d-2) C^d \frac{z_0^{d-3}}{(z 
- \vec{x})^{2 (d - 1)}} \,I_{0[ \mu}(z-\vec{x})I_{\nu ]i}(z -
\vec{x})\,,
\end{eqnarray}
where $\partial_{[\mu}G^G_{\nu ]i}$ is gauge invariant. 
We note that if the second term in (\ref{covidentity}) had the factor 
$(d-1)^{-1}$ instead of $d^{-1}$, $\nabla \nabla G$ would have the index
symmetry of the Weyl tensor in five dimensions when $d=4$. However since the
factor is $d^{-1}$ this seems not to be the case. Although the vertex
(\ref{genexpand}) has Weyl index symmetry in $d = 10$, this structure 
seems to be lost when just considering the AdS part of the
compactification $AdS_5 \times S_5$.  See also \cite{Tseytlin:2000sf}. 

With these ingredients the stress tensor four-point function defined in 
(\ref{genexpand}) is given by
\begin{eqnarray}
\lefteqn{T_{a_1 b_1 a_2 b_2 a_3 b_3 a_4 b_4} 
 (\vec{x}_1,\vec{x}_2,\vec{x}_3,\vec{x}_4)}  & & \nonumber \\ 
&=&
\int\!\!\frac{d^{d+1}z}{z_0^{d+1}} t^{8 \mu_1 \sigma_1 \cdots \mu_4
\sigma_4}t^{8 \nu_1 \rho_1 \cdots \nu_4 \rho_4}
 \nabla_{[\rho_1} \nabla_{[\sigma_1} G_{\mu_1 ] \nu_1
],a_1 b_1}(z-\vec{x}_1)\cdots\nabla_{[\rho_4} \nabla_{[\sigma_4} G_{\mu_4 ] \nu_4
],a_4 b_4}(z-\vec{x}_4) \nonumber \\ 
&=& 
\int\!\!\frac{d^{d+1}z}{z_0^{d+1}} t^{8}_{ \mu_1 \sigma_1 \cdots \mu_4
\sigma_4}t^{8}_{\nu_1 \rho_1 \cdots \nu_4 \rho_4}
\E_{a_1 b_1, k_1 l_1}\cdots \E_{a_4 b_4, k_4 l_4} \label{generalform}\\
& & \; \qquad\quad \times \,
\I^R_{0 \mu_1 \sigma_1 k_1 l_1 \rho_1 \nu_1 0} (z - \vec{x}_1) 
\I^R_{0 \mu_2 \sigma_2 k_1 l_2 \rho_2 \nu_2 0} (z - \vec{x}_2) \nonumber\\
& & \; \qquad\quad \times \, 
\I^R_{0 \mu_3 \sigma_3 k_1 l_3 \rho_3 \nu_3 0} (z - \vec{x}_3) 
\I^R_{0 \mu_4 \sigma_4 k_1 l_4 \rho_4 \nu_4 0} (z - \vec{x}_4)\,, \nonumber
\end{eqnarray}
where 
\begin{eqnarray}
 \I^R_{0 \mu \sigma k l \rho \nu 0} (z - \vec{x}) &=& 
\left[\frac{\Gamma(d+2)}{\pi^{d/2}\Gamma(d/2)}\right]
\frac{z_0^d}{(z-\vec{x})^{2d}}\,( I_{0[ \mu} I_{\sigma ] k} I_{ l [
\rho }I_{ \nu ] 0 } - {\ts \frac{1}{d}} \delta_{[ \mu [ \nu }I_{\rho ]
k } I_{ \sigma ] l } ) (z - \vec{x})\,.
\end{eqnarray} 
The overall factor of $z_0^{16}$ compensates for the lowering of sixteen  
indices.  For reference the explicit expression for the $t^8$ tensor is 
given in the Appendix.  

We perform an inversion in $\vec{x}_1$ according to the method of 
\cite{Freedman:1999tz} and define
\begin{equation}
\vec{x}'_{21\,k} = \frac{ \vec{x}_{21\,k}}{ |\vec{x}_{21}|^2} \, .
\end{equation}
This gives for the four-point correlation function as given by (\ref{generalform})
\begin{eqnarray}
\lefteqn{T_{a_1 b_1 a_2 b_2 a_3 b_3 a_4
b_4}(\vec{x}_1,\vec{x}_2,\vec{x}_3,\vec{x}_4)}  & & \nonumber \\ 
&=&
\frac{\I^T_{a_2 b_2, k_2 l_2} (\vec{x}_{21})
\I^T_{a_3 b_3, k_3 l_3} (\vec{x}_{31})
\I^T_{a_4 b_4, k_4 l_4} (\vec{x}_{41}) \E_{a_1 b_1 k_1 l_1}}
{ |\vec{x}_{21}|^{2d} 
  |\vec{x}_{31}|^{2d} 
  |\vec{x}_{41}|^{2d} }  t^{TTTT}_{k_1 l_1k_2 l_2k_3 l_3k_4 l_4} (
\x'_{23}, \x'_{24})\, , \label{klaus1} 
\end{eqnarray} 
where 
\begin{eqnarray} 
\lefteqn{t^{TTTT}_{k_1 l_1k_2 l_2k_3 l_3k_4 l_4} (
\x'_{23}, \x'_{24})} & & \nonumber\\ && 
= \int\!\!\frac{d^{d+1}z'}{z'_0{}^{d+1}} t^{8}_{ \mu_1 \sigma_1 \cdots \mu_4
\sigma_4}t^{8}_{\nu_1 \rho_1 \cdots \nu_4 \rho_4} 
(\delta_{ 0 [ \mu_1 }\delta_{ \sigma_1 ] k_1 } \delta_{ l_1 [ \rho_1 } 
\delta_{ \nu_1 ] 0 } - {\ts \frac{1}{d}} \delta_{ \mu_1 [ \nu_1}
\delta_{ \rho_1 ] k_1} \delta_{\sigma_1 ] l_1 })
\nonumber\\ && \;\; \times
\I^R_{0 \mu_2 \sigma_2 k_2 l_2 \rho_2 \nu_2 0} (z' - \vec{x}'_{21}) 
\I^R_{0 \mu_3 \sigma_3 k_3 l_3 \rho_3 \nu_3 0} (z' - \vec{x}'_{31}) 
\I^R_{0 \mu_4 \sigma_4 k_4 l_4 \rho_4 \nu_4 0} (z' - \vec{x}'_{41})\,.
\label{klaus}
\end{eqnarray}
$\I^T$ is given by (\ref{it}).
In defining $t^{TTTT}$ we have used the fact that the integral in (\ref{klaus})
depends only on the differences
\begin{equation} 
\label{k2}
\vec{x}'_{23\,k}=\frac{\vec{x}_{21\,k}}{|\vec{x}_{21}|^2} - 
\frac{\vec{x}_{31\,k}}{|\vec{x}_{31}|^2} \, , \quad
\vec{x}'_{24\,k}=\frac{\vec{x}_{21\,k}}{|\vec{x}_{21}|^2} - 
\frac{\vec{x}_{41\,k}}{|\vec{x}_{41}|^2} \, .
\end{equation}
Note that these variables are of the same form as the variable $X$ in (\ref{it}).
The expression (\ref{klaus1}) is very reminiscent of the general construction
procedure for conformal three-point functions of \cite{Osborn:1994cr,Erdmenger:1997yc} 
as described in \ref{OPEsec}.1, 
and gives some indication how this procedure may be generalized 
to four-point functions. 

The next step is the evaluation of the integral in (\ref{klaus}).
Since this is a tedious task we just outline the calculation here
and proceed to the determination of the leading terms when
$\vec{x}_1 \rightarrow \vec{x}_2$.  In the general case the evaluation of
the integral necessitates the extraction of derivatives with respect 
to the $\vec{x}'_{m}$ such as to obtain a collection of scalar integrals.
In order to achieve this, we decompose the sum over indices in 
(\ref{klaus}) by writing $\mu\,\hat{=}\, (0,m)$, where $0$ stands for the
bulk coordinate and the Latin indices for the coordinates parallel to the
boundary. We demonstrate this procedure using the simple example of two 
antisymmetric tensors $M$ and $N$:
\begin{equation}
M_{\mu\nu}N_{\mu\nu} = 2 M_{0 n} N_{0 n} + M_{m n }N_{m n}.
\end{equation}
Some terms in the sum in (\ref{klaus}) obtained this way vanish due to 
\begin{equation}
t^8_{0 k i j m n p q} = 0\quad t^8_{0 k 0 j 0 n p q} = 0.
\end{equation}
From the expressions 
\begin{equation}
\I^R_{0 m s k l r n 0} \,, \I^R_{0 0 s k l r 0 0 }\,,
\I^R_{0 m s k l r 0 0} 
\end{equation}
the necessary derivatives with respect to $\vec{x}$ may be extracted
using the algebraic computing program FORM. The results are given in 
the Appendix. 

The procedure outlined here gives a sum of scalar integrals of the
type
\begin{equation}
\I =
\int\!\!\frac{d^{d+1}z'}{z'_0{}^{d+1}}\,\frac{z_0^{\sum\limits_{i=1}^{4}\Delta_i}}
{(z'-\vec{x}'_{21})^{2\Delta_2}(z'-\vec{x}'_{31})^{2\Delta_3}(z'-\vec{x}'_{41}
)^{2\Delta_4}}\,, 
\label{fritz}
\end{equation}
with a collection of derivatives acting on them. There are at most
four derivatives for every $x$. The tensor structure is completed by a 
number of Kronecker $\delta$'s. The integral (\ref{fritz}) may be
evaluated using Feynman parameter methods which give 
\begin{eqnarray} 
\label{Feynman}
\I &=& \pi^{d/2}\frac{\Gamma[(\sum\limits_{i=1}^4 \Delta_i -d)/2] \Gamma 
[ \half(\Delta_2+\Delta_3+\Delta_4 -
\Delta_1)]}{2\Gamma[\Delta_2]\Gamma[\Delta_3]\Gamma[\Delta_4]}\\
& & \times\,\int\!\!\ \prod_{j=2}^4 d\alpha_j ~ 
\delta(1-\sum\limits_i \alpha_i) ~ 
 \frac{{\alpha_j^{\Delta_j - 1}}}{(\alpha_2 \alpha_3 |\vec{x}'_{23}|^2 
+ \alpha_2 \alpha_4 |\vec{x}'_{24}|^2 + \alpha_3 \alpha_4
|\vec{x}'_{34}|^2 )^{\half(\Delta_2 +\Delta_3 + \Delta_4 - \Delta_1)}} \,, 
\nonumber 
\label{parameterintegral}
\end{eqnarray}
with $\x'_{23}, \x'_{24}$ given by (\ref{k2}) and $\x'_{34}=\x'_{24}- \x'_{23}$.
By a standard change of variables we obtain 
\begin{eqnarray}
\I &=& \pi^{d/2}\frac{\Gamma[(\sum\limits_{i=1}^4 \Delta_i -d)/2] \Gamma 
[ \half(\Delta_2+\Delta_3+\Delta_4 - \Delta_1)]} 
 {2\Gamma[\Delta_2]\Gamma[\Delta_3]\Gamma[\Delta_4]}\\ \nonumber
& & \times\,\int\!\!
d {\beta_2}d {\beta_3}\frac{\beta_2^{\Delta_2-1}\beta_3^{\Delta_3 - 1}} 
 { (1 + \beta_2 + \beta_3)^{\Delta_1}}\,\frac{1}{|\vec{x}'_{23}|^{\half
(\Delta_2+\Delta_3+\Delta_4-\Delta_1)} (\beta_2 \beta_3 + \xi\beta_3
+\eta\beta_2)^{\half(\Delta_2+\Delta_3+\Delta_4-\Delta_1)}} \ ,  
\label{iintegral} 
\end{eqnarray}
where we have introduced the conformally invariant cross-ratios, 
\begin{equation}
\eta = \frac{|\x'_{24}|^2}{|\x'_{23}|^2}=\frac{|\x_{24}|^2
|\x_{31}|^2}{|\x_{23}|^2 |\x_{41}|^2}\,,\quad 
\xi  = \frac{|\x'_{34}|^2}{|\x'_{23}|^2}=\frac{|\x_{34}|^2
|\x_{21}|^2}{|\x_{23}|^2 |\x_{41}|^2}\, . 
\label{crossratios} 
\end{equation}
In general this integral gives a power series involving a generalized 
hypergeometric function.  However for our purpose it is sufficient to consider 
the case when $|\x_{12}|\rightarrow 0$ in which 
\begin{equation}
\eta \rightarrow 1 \,,\quad \xi \rightarrow |\x_{21}| ^2
\frac{|\x_{34}|^2}{|\x_{32}|^2 |\x_{34}|^2}\, = |\x_{21}| ^2 |\x'_{34}| ^2 \, ,
\end{equation}
and $|\x'_{23}| ^{-\Delta} \simeq |\x_{21}| ^\Delta$.  Although the form in 
\rf{iintegral} is sufficient to extract the leading behavior, we perform 
a further transformation as in \cite{Brodie} to simplify the analysis, 
\begin{equation}
\beta_2 \rightarrow \beta_2' = \frac{\beta_2}{\beta_3}(1+\beta_3)\,,
\end{equation}
which gives, omitting the overall prefactor,  
\begin{eqnarray}
\I' &=& \int\!\! d \beta_2' d \beta_3\,\frac{\beta_3^{\half(\Delta_1
+ \Delta_2 +\Delta_3 - \Delta_4)}}{(1+\beta_3)^{\Delta_2} (1+\beta_3
+\beta_2' \frac{\beta_3}{1+\beta_3})^{\Delta_1}} \, 
\frac{\beta'_2{}^{\Delta_2 - 1}}{(\beta'_2+\xi)^{\half
(\Delta_2 + \Delta_3 +\Delta_4 -\Delta_1)}}\,. 
\label{susi}
\end{eqnarray}
For $\Delta_2+\Delta_1 > \Delta_3 + \Delta_4$ we see that when $\xi
\rightarrow 0$ the dominant contribution to the $\beta_2'$
integral comes from the region where $\beta'_2 \simeq 0$. In this
region the second tern in the denominator of (\ref{susi}) is independent 
of $\beta'_2$ and the integral factors to give 
\begin{eqnarray} 
\I &\sim & K \, |\x_{21}| ^{\Delta_2+ \Delta_3 + \Delta_4 -\Delta_1} \,  
\xi^{\half (\Delta_2+\Delta_1 -\Delta_3 -\Delta_4)}\, 
= \, K |\x_{21}| ^{2\Delta_2} 
|\x'_{34}|^{ (\Delta_2+\Delta_1 -\Delta_3 -\Delta_4)}\,, \label{x12} \\
K &=& \frac{1}{\Gamma[\Delta_1+ \Delta_2] \Gamma[\Delta_3]  \Gamma[\Delta_4]}\, 
\Big\{\Gamma[(\sum \Delta_i - d)/2]\Gamma[(\Delta_3 +\Delta_4 - \Delta_2 
- \Delta_1)/2] \nonumber\\ &&  
\hskip -.1cm \times \, 
\Gamma[(\Delta_3 +\Delta_4 - \Delta_2 
- \Delta_1)/2]
\Gamma[(\Delta_1 + \Delta_2 + \Delta_3 - \Delta_4 )/2 ]
\Gamma[(\Delta_1 + \Delta_2 + \Delta_4 - \Delta_3 )/2 ]\Big\}
\,.\nonumber
\end{eqnarray}
For $\Delta_2+\Delta_1 = \Delta_3 + \Delta_4$ we perform the $\beta'_2$ integral 
explicitly which leads to a hypergeometric function whose leading term involves
${\rm ln} \xi$. In this case we have
\begin{equation}
\I \sim \frac{\pi^d}{2} \frac{\Gamma[(\sum \Delta_i -d)/2]} 
 {\Gamma[\Delta_1+ \Delta_2]}
\, |\x_{21}|^{2 \Delta_2} \, ( {\rm const.} - {\rm ln} \xi) \, .
\end{equation}

With the help of these general results we may determine the leading behaviour of
(\ref{klaus}). As an example for a term present in (\ref{klaus}) we consider
\begin{equation}
\delta _{ a_1 b_1} \delta _{a_2 b_2 } \delta_{a_3 b_3} \delta_{a_4 b_4}
P_0 \equiv \delta_{a_1 b_1} \delta_{a_2 b_2} \delta_{a_3 b_3} \delta_{a_4 b_4}
\int \!\!  \frac{d ^{ d+1} z'} {z'_0 {} ^ {d+1} }  \;
\frac{ z'_0 {} ^ d } { (z'- \x'_{21}) ^{ 2d} (z'- \x'_ {31}) ^{2d}
(z'- \x'_ {41}) ^{2d}}  \,
\label{4delta}
\end{equation}

Here no extraction of derivatives is necessary since the tensor structure is 
entirely given by Kronecker deltas. Using the results for Feynman integrals 
described above, in particular (\ref{x12}), we find for $ \x_1 \rightarrow \x_2 $:
\begin{equation}
P_0 \propto    |\x_{21}|^ {2d} \,  ( {\rm const.} -  \ln \xi ) \quad . 
\label{particular}
\end{equation}
Together with the factor of $  \I ^T / x_{21} ^{2d} $ present in (\ref{klaus}) 
we see that in the limit $ \x_1 \rightarrow \x_2 $ the contribution 
(\ref{4delta}) depends only on dimensionless combinations of $ \x_{21} $ 
such as $ \x_{21\, a} \x_{21\,b} / x_{21} {}^2 $ and $\ln\xi$.  (\ref{4delta}) 
does not give rise to any poles for which according to (\ref{limit2}) a 
dependence of the form $|\x_{21}| ^{- \lambda}$ with $\lambda \geq d$ 
is necessary.

By simple power-counting we may determine the $ |\x_{21}|$ 
dependence of the terms in (\ref{klaus}) involving derivatives. 
In extracting the derivatives a dimensional regulation in $d$ is used 
to ensure finite scalar integrals at every step of the calculation.  
However after taking the derivatives again after evaluating the integrals, 
all expressions are finite such that the regulator may be taken to zero. 

By inspection of the expressions for the $\I^R$ listed in the appendix we 
see that the derivative terms are of the general form
\begin{eqnarray} 
P_\pr &=& \delta \cdots \delta \, \pr^{x'_{21}} \cdots
\pr^{x'_{21}}\, \pr^{x'_{31}} \cdots \pr^{x'_{31}}\,
\pr^{x'_{41}} \, \cdots \pr^{x'_{41}} \; \; \nonumber\\ && \hspace{0.1cm} 
\times \, 
\int \! \! \frac{d^{d+1}z'} {z_0^{d+1}} \,
\frac{z_0'{}^{4d+2 \sum a_i+b}}
        {(z'-\x_{21})^ {2(d-n_2 /2+a_2)}
        (z'- \x_{31})^{2(d-n_3/2+a_3}
        (z'- \x_{41})^{2(d-n_4/2+a_4)} } \, ,
\label{ppr}
\end{eqnarray}
where the tensor structure ($\delta \cdots \delta)$ in \rf{ppr} is determined 
by a combination of Kronecker deltas and derivatives. Furthermore defining 
$q_i$ to be the total number of derivatives extracted with respect to the 
variable $x'_{i1}$, we have for the numbers $n_2, n_3, n_4$
in (\ref{ppr})
\begin{eqnarray*}
q_i ~ \, {\rm even} \, &:& n_i \equiv q_i \\
q_i ~ \, {\rm odd } \, &:& n_i \equiv q_i +1 \, .
\end{eqnarray*}
The $a_i$ are integers for which $a_i \geq 0$. $b$ is an integer for which $2 
\geq b \geq 0$, which is non-zero if an odd number of derivatives is extracted 
for any of the three variables ${\vec x}'_{21}, {\vec x}'_{31}, {\vec x}'_{41}$. 
For the discussion here we note that \rf{ppr} may be written as a Feynman 
parameter integral of the form \rf{parameterintegral}.  These integrals 
are evaluated explicitly in Section 5.3.

We now consider the expression we obtain when consecutively evaluating
the derivatives explicitly when acting on the Feynman parameter representation 
\rf{parameterintegral} and \rf{ppr}.  A given derivative $\pr_p^{x'_{ij}}$ may 
act on the denominator of the Feynman parameter integral or on factors $x'_{ij}$ 
($i,j\in \{2,3,4\}$) in the numerator created by the action of derivatives 
evaluated {\it before} $\pr_p^{x'_{ij}}$.

For determining the scaling behaviour of the $\x_{12}$ dependence it is
convenient to define the following integers:
\begin{eqnarray*}
n_{ij}^{(i,D)}  &: & {\rm number\, of\, derivatives\,  with\,  respect\,
 to} \,  x'_{i1} \, {\rm acting \, 
on} \, x'_{ij} \,  {\rm in \, the\, denominator} \\ &&
 {\rm \, of\, the\, Feynman\, parameter\, integral} \\
n_{ij}^{(i,N)} &:& {\rm number \, of\, derivatives\, with\, respect\, to} \,
 x'_{i1} \, {\rm acting \, 
on} \,  x'_{ij} \, {\rm in \,  the \, numerator}\\
N_N^{(2)} &\equiv &
 n_{23}^{(2,N)}+n_{24}^{(2,N)}+n_{23}^{(3,N)}+n_{24}^{(4,N)} \, , \\
N_D^{(2)} &\equiv &
n_{23}^{(2,D)}+n_{24}^{(2,D)}+n_{23}^{(3,D)}+n_{24}^{(4,D)}  .
\end{eqnarray*}
We have
\begin{eqnarray}
n_2 &=& n_{23}^{(2,N)}+n_{2,4}^{(2,N)}+n_{23}^{(2,D)}+n_{24}^{(2,D)}\, 
\qquad {\rm for} \, q_2 ~ \, {\rm even} \, , \nonumber\\
n_2  &=& n_{23}^{(2,N)}+n_{2,4}^{(2,N)}+n_{23}^{(2,D)}+n_{24}^{(2,D)} + 1\, 
\quad {\rm for} \, q_2 ~ \, {\rm odd} \, . \nonumber\\
\label{n2}
\end{eqnarray}
With these definitions we obtain
\begin{eqnarray}
P_{\pr} &=& ( x'_{23,l_2} \cdots x'_{34,l_{(D-N)}})(\delta \cdots \delta
)\int 
\,\!\!\ \prod_{j=2}^4 d\alpha_j ~ \delta(1-\sum\limits_i \alpha_i) ~ 
\nonumber\\ 
& & 
\times 
 \frac{{\alpha_j^{\Delta_j - 1}}}{(\alpha_2 \alpha_3 |\vec{x}'_{23}|^2 
+ \alpha_2 \alpha_4 |\vec{x}'_{24}|^2 + \alpha_3 \alpha_4
|\vec{x}'_{34}|^2 )^{\half(\Delta_2 +\Delta_3 + \Delta_4 - \Delta_1)}} \,.
\label{p2}
\end{eqnarray}
where 
\begin{equation}\label{delta2}
\Delta_2 = d - n_2/2 + N_D^{(2)}+a_2 \, , \quad
\Delta_1 = d + \sum n_i + \sum a_i + b . 
\end{equation}
$\Delta_3,\Delta_4$ are given by expressions similar to $\Delta_2$.
We always have $\Delta_1+ \Delta_2 \geq \Delta_3+ \Delta_4$. 
The first factor in (\ref{p2}) 
is a tensor product of vectors of which $N_D^{(2)}-N_N^{(2)}$ contain $\x'_{21,l}$.  
In the limit $\x_1 \to \x_2$ we calculate the integral using (\ref{x12}) 
or \rf{particular} to obtain
\begin{equation}
P_{\partial}\propto \left( \frac{\x_{21,l_1}}{|\x_{21}|^2}\cdots
  \frac{\x_{21,l_{(D-N)}}}{|\x_{21}|^2}\right) \, |\x_{21}|^{2\Delta_2} \, 
  ({\rm const.} - \ln \xi)  \, t(\x'_{34})\quad . \label{pfinal}
\end{equation}
Here $\x'_{34}$ is now independent of $\x_1$, 
\begin{equation}
\x'_{34,l} = \frac{\x_{32,l}}{|\x_{32}|^2} -
  \frac{\x_{42,l}}{|\x_{42}|^2}\quad .
\end{equation}
$t$ is a tensor which is of a similar structure as the tensors $t$ in
(\ref{TTI}).
In (\ref{pfinal}) there are $N_D^{(2)} - N_N^{(2)}$ factors of
  $\x_{21}/|\x_{21}|^2$.
Remembering the factor $\I^T/|\x_{21}|^{2d}$ in (\ref{klaus}) we see that
all the terms contributing to (\ref{klaus}) depend at most on dimensionless 
combinations of $\x_{21}$ if 
\begin{equation} \label{cond}
2\Delta_2 - N_D^{(2)} + N_N^{(2)} \geq 2d \quad ,
\end{equation}
which using the definitions of the integers defined in the table
above (\ref{n2}) as well as (\ref{delta2})  is equivalent to 
\begin{equation}
2a_2 + n_{24}^{(4,D)} + n_{23}^{(3,D)} + n_{24}^{(4,N)} + n_{23}^{(3,N)}
\geq 0 \, .
\end{equation}
Since by definition all the integers in this condition are non-negative,
this condition is always satisfied. Therefore we have proved that the
contribution to the stress-tensor four-point function given by
(\ref{klaus}) contributes only non-leading terms involving
operators of dimension $\eta \geq  2d$ to the stress-tensor OPE.

It is straightforward to see that our argument extends also to contributions
with $k \neq 0$ in (\ref{genexpand}). Any additional $\Box^z= g^{\mu \nu} 
\nabla^z_\mu \nabla^z_\nu$ increases $N_D^{(2)}$ by $2$ at most, but
increases $\Delta_2$ by the same amount at the same time such that 
(\ref{cond}) is not modified.  Moreover the general terms in the
string scattering elements, including those involving the background values 
of the five-form field strength in \rf{fiveform}, are of the form $\Box^k R^4$, 
and in a similar way these terms do not contribute any leading terms to the 
OPE through the holographic scattering.

Finally we note that the absence of leading terms in the OPE implies that the
contributions to the stress-tensor four-point function discussed here are not 
linked to the stress-tensor two- and three-point function by any Ward identity,
which is consistent with the non-renormalization properties of the two- 
and three-point functions. 
In fact, the diffeomorphism and Weyl symmetry Ward identities
expressing conservation and tracelessness of the stress tensor are of the
form
\begin{eqnarray}
\pr^{\x_1}_{a_1} \langle
T_{a_1 b_1}(\x_1)T_{a_2 b_2}(\x_2)T_{a_3 b_3}(\x_3)T_{a_4 b_4}(\x_4) \rangle
&=&\pr_{b_1} \delta^{(d)} (\x_{12}) 
\langle T_{a_2 b_2}(\x_2)T_{a_3 b_3}(\x_3)T_{a_4 b_4}(\x_4) \rangle
+ \dots \, ,\nonumber\\
\langle
T_{a_1 a_1}(\x_1)T_{a_2 b_2}(\x_2)T_{a_3 b_3}(\x_3)T_{a_4 b_4}(\x_4) \rangle
&=& \delta^{(d)} (\x_{12}) 
\langle T_{a_2 b_2}(\x_2)T_{a_3 b_3}(\x_3)T_{a_4 b_4}(\x_4) \rangle
+ \dots \, ,
\end{eqnarray}
where the dots denote permutations of $(2 \leftrightarrow 3 \leftrightarrow
4)$ and terms involving two-point functions. 
According to (\ref{limit2}), the necessary delta distributions on the right
hand side of the Ward identity are generated by those terms on the left 
hand side which in the limit $\x_1 \rightarrow \x_2$
involve $|\x_{21}|^{-\lambda}$ with $\lambda \geq d$, and similarly for
$\x_3, \x_4$. As we have shown above, the four-point function contributions
discussed here do not involve such terms. 

However we expect the stress-tensor four-point function contribution
corresponding to exchange diagrams in the supergravity approximation
to contain such leading terms, such that the Ward identities above 
are satisfied by the full stress-tensor four-point function obtained
from the AdS/CFT correspondence.  Moreover we expect
massless supergravity exchange
contributions to contain terms which factor into two-point functions (at 
leading order in $\lambda$ large).

\subsection{Triangle diagram structure}

In this section we give the form of all of the triangle integrals 
necessary to reproduce the contributions to the four-point correlation 
function at intermediate values of the coupling within our approach,
except for the non-analytic terms which are not considered here.  The form 
of the integral in \rf{fritz} is related to a triple off-shell triangle 
integral with massless internal lines.   The $d$ dimensional
version of which has been evaluated in terms of Appel functions in
\cite{Anastasiou:2000ui,Boos:1991rg}.  We give the form here, together with 
the parameters necessary to map between the two expressions.  

The off-shell triangle described by 
\bqr  
I[p_1,p_2,p_3] = \int {d^dl\over (2\pi)^d} ~ {1\over l^{2d_1} 
 (l-p_1)^{2d_2} (l-p_1-p_3)^{2d_3} } 
\label{offshelltriangle} 
\fqr   
is related to the integral in \rf{fritz} by the redefinition of the 
variables, 
\bqr  
p_1 = {\vec x}_{32} \qquad p_3 = {\vec x}_{24}  
\ , 
\fqr 
and in \rf{p2} with 
\bqr  
Q_1^2 = |{\vec x}_{23}|^2 \ , \,  Q_2^2 = |{\vec x}_{24}|^2 \ , \,   
 Q_3^2 = |{\vec x}_{34}|^2 \ , 
\label{Qidentify}  
\fqr 
together with the identification of dimension $d$ and powers $d_i$ 
\bqr  
d= \sum_{j=1}^4 \Delta_j ,\qquad d_1 = \Delta_2 , \quad 
d_2 = \Delta_3 , \quad d_3 = \Delta_4 \ ,   
\fqr 
in addition to a normalization factor found after comparing the 
two integrals in a Schwinger proper time form.   In comparing the integral 
forms we appeal directly to the latter form in \rf{p2} and refer the 
reader to \cite{Anastasiou:2000ui,Boos:1991rg} for the derivation.  

The result for the integral in \rf{fritz} in terms of the two 
independent conformal invariant cross-ratios in \rf{crossratios} (where 
$\eta=Q^2_2/Q_1^2$ and $\xi=Q_3^2/Q_1^2$) is, 
\bqr 
I_{\Delta_j}(\eta,\xi) = \sum_{j=1}^4 A^{(j)}_{\Delta_j}(\eta,\xi) \ , 
\label{triangles} 
\fqr 
with  
\bqr \hskip -.5cm  
A^{(1)}_{\Delta_j}(\eta,\xi) = (-1)^{d/2} \Bigl({\vec x}_{23}^2 
 \Bigr)^{{d\over 2}-\Delta_2 - \Delta_3  - \Delta_4} 
{\Gamma\left({d\over 2}-\Delta_2-\Delta_3\right) \Gamma\left({d\over 2}- 
 \Delta_2-\Delta_4\right) \Gamma\left(\Delta_2+\Delta_3 - {d\over 2}\right) 
\over \Gamma\left(\Delta_3\right) \Gamma\left(\Delta_4\right) 
 \Gamma\left(d-\Delta_2-\Delta_3\right)} \nonumber 
\fqr 
\bqr \times 
F_4\Bigl( \Delta_2, \Delta_2+\Delta_3-{d\over 2}, 1+\Delta_2+\Delta_4 - {d\over 2}, 
 1+\Delta_2+\Delta_3-{d\over 2}; \eta,\xi \Bigr) \ , 
\fqr 
\bqr  \hskip -.5cm
A^{(2)}_{\Delta_j}(\eta,\xi) = (-1)^{d/2} \Bigl({\vec x}_{23}^2\Bigr)^{-\Delta_3} 
\Bigl({\vec x}_{24}^2\Bigr)^{{d\over 2}-\Delta_2-\Delta_4} 
{\Gamma\left({d\over 2}-\Delta_2-\Delta_3\right) \Gamma\left(\Delta_2+\Delta_4 
-{d\over 2}\right) \Gamma\left({d\over 2}-\Delta_4\right)\over \Gamma(\Delta_2) 
\Gamma\left(\Delta_4\right) \Gamma\left(d-\Delta_2-\Delta_3\right) }
\nonumber 
\fqr 
\bqr \times 
F_4\Bigl( \Delta_3, {d\over 2}-\Delta_4, 1+ {d\over 2}-\Delta_2-\Delta_4, 
 1+ \Delta_2+\Delta_3 -{d\over 2}; \eta,\xi\Bigr) \ , 
\fqr 
\bqr  \hskip -.5cm
A^{(3)}_{\Delta_j}(\eta,\xi) = (-1)^{d/2} \Bigl({\vec x}_{23}^2\Bigr)^{-\Delta_4} 
\Bigl({\vec x}_{24}^2\Bigr)^{{d\over 2}-\Delta_2-\Delta_3} { 
\Gamma\left({d\over 2}-\Delta_2-\Delta_4\right) \Gamma\left(\Delta_2+\Delta_3-
 {d\over 2}\right) \Gamma\left({d\over 2}-\Delta_3\right)\over 
\Gamma\left(\Delta_2\right) \Gamma\left(\Delta_3\right) \Gamma\left(d-\Delta_2- 
\Delta_3\right)} \nonumber 
\fqr 
\bqr  \times 
F_4 \Bigl( \Delta_4, {d\over 2}-\Delta_3, 1+\Delta_2+\Delta_4-{d\over 2}, 
 1+{d\over 2}-\Delta_2-\Delta_3; \eta,\xi \Bigr) \ , 
\fqr 
and, 
\bqr  
\hskip -2cm
A^{(4)}_{\Delta_j}(\eta,\xi) = (-1)^{d/2} \Bigl({\vec x}_{23}^2\Bigr)^{\Delta_2 
- {d\over 2}} \Bigl({\vec x}_{24}^2\Bigr)^{{d\over 2}-\Delta_2-\Delta_4} 
\Bigl({\vec x}_{34}^2\Bigr)^{{d\over 2}-\Delta_2-\Delta_3} 
\nonumber 
\fqr 
\bqr 
\times 
{\Gamma\left(\Delta_2+\Delta_3-{d\over 2}\right) \Gamma\left( \Delta_2 + 
\Delta_4-{d\over 2}\right) \Gamma\left( {d\over 2}-\Delta_2\right) \over 
\Gamma\left( \Delta_2\right) \Gamma\left(\Delta_3\right) \Gamma\left( 
\Delta_4\right)} \nonumber 
\fqr 
\bqr \times 
F_4\Bigl( d-\Delta_2-\Delta_3, {d\over 2}-\Delta_2, 1+{d\over 2} - 
\Delta_2-\Delta_4, 1+{d\over 2}-\Delta_2-\Delta_3; \eta,\xi \Bigr) \ .  
\fqr 
These Appel functions are defined by the double infinite series expansion, 
\bqr  
F_4(\alpha,\beta,\gamma,\delta;x,y) = 
\sum_{(m,n)}^{\infty} {\Gamma\left(\alpha+m+n\right)\Gamma\left(\beta 
+m+n\right) \Gamma\left(m\right) \Gamma\left(n\right)\over 
\Gamma^2\left(m+n\right) \Gamma(\gamma+m)\Gamma\left(\delta+n\right)} 
{x^m\over \Gamma(m+1)} {y^n\over \Gamma\left(n+1\right)} \ ,  
\fqr 
and their integral representations, from which an analytic continuation 
may obtained.  We refer the reader to \cite{Anastasiou:2000ui,Boos:1991rg} 
or references therein for further details.  The forms above for the integrals 
are useful for finding the complete OPE of the $\N=4$ super Yang-Mills gauge 
theory away from infinite coupling. 

\subsection{Comments on box diagram structure and $n\geq 4$ points} 

An alternative representation for the contributions to the stress-tensor
four-point function given by (\ref{generalform}) may be found without 
performing an inversion which leads to a box diagram structure. We find
by expanding the inversion tensor $I_{ab}$ and extracting an explicit 
$(z-{\vec x}_j)^{-4}$ from each graviton propagator in (\ref{generalform})
\bqr  
\int {dz_0 d^d{\vec z}\over z_0^{d+1}} 
 \prod_{j=1}^4 {z_0^d \over (z-{\vec x}_j)^{2d+4}}  T_{\{\mu_j \nu_j\} 
 \{{\bar\mu}_j{\bar\nu}_j\} \{a_j b_j\}} \ ,  
\label{boxstructure}
\fqr  
where the $\mu_j$ and $\nu_j$ indices on $T_{\{\mu_j \nu_j\} 
 \{{\bar\mu}_j{\bar\nu}_j\} \{a_j b_j\}}$ are to be contracted with the 
two $t_8$ tensors.  The result for $T$ is 
\bqr  
T &=& \prod_{j=1}^4 \Bigl\{ ~\left[ 
 (z-{\vec x}_j)^2 \delta_{0[\mu_j} \delta_{\nu_j]a_j} 
 -2 z_0 (z-{\vec x}_j)_{[\mu_j} \delta_{\nu_j]a_j} - 2 
 (z-{\vec x}_j)_{[\mu_j\vert} (z-{\vec x}_j)_{a_j} \delta_{\vert\nu_j]0} \right]  
\non && \hskip .22in
\times \left[ 
 (z-{\vec x}_j)^2 \delta_{0[{\bar\mu}_j} \delta_{{\bar\nu}_j]b_j} 
 -2 z_0 (z-{\vec x}_j)_{[{\bar\mu}_j} \delta_{{\bar\nu}_j]b_j} - 2 
 (z-{\vec x}_j)_{[{\bar\mu}_j\vert} (z-{\vec x}_j)_{b_j} 
 \delta_{\vert{\bar\nu}_j]0} \right]  
\non && \hskip 2in
+ (a_j \leftrightarrow b_j) 
\non &&  \hskip -.37in
-{1\over d} \delta_{a_j b_j} \left[ (z-{\vec x}_j)^2 \delta_{0[\mu_j\vert} 
 -2 z_0 (z-{\vec x}_j)_{[\mu_j\vert} \right] 
 \left[ (z-{\vec x}_j)^2 \delta_{0[{\bar\mu}_j} 
 -2 z_0 (z-{\vec x}_j)_{[{\bar\mu}_j} \right] \delta_{\vert\nu_j]{\bar\nu}_j]} ~ 
\Bigr\} \ . 
\label{tform}
\fqr  
\rf{boxstructure} 
is of the form of a field theory box 
diagram as calculated below, and the expansion of \rf{tform} modifies 
it with derivatives or different powers of $(z-{\vec x}_j)^2$.  The 
general term found from expanding $T$ in \rf{tform} is constructed by 
differentiating the integral
\bqr  
\cI_{a;a_j} = \int {dz_0 d^d{\vec z}\over z_0^{d+1}} ~ z_0^{a} ~
\prod_{j=1}^4 {z_0^d \over (z-{\vec x}_j)^{2d+2a_j} } \ , 
\label{Ifunction}
\fqr 
with $a\leq 16$ and $a_j\leq 2$.  An analysis of the singularities within 
the OPE structure of \rf{Ifunction} may be performed on the box integrals, 
but we have analyzed this in the reduced inverted form on the triangle 
representation.  Conformal invariance allows the representation in \rf{tform} 
to be simplified into a triangle form (together with external derivatives).  

We note that for more general correlator examples at $n$-point 
the relation between the $n$-point scalar integral functions 
persists with that of the holographic Feynman-Witten diagrams.  
For example, the scalar holographic integral at $n$-point is 
described by 
\bqr  
I_{a;a_j}^\gamma = \int {dz_0 d^d{\vec z}\over z_0^{d+1}} ~z_0^\gamma ~
\prod_{j=1}^n {z_0^{\Delta_j} \over (z-{\vec x}_i)^{2\Delta_j}} \ , 
\label{npointhol}
\fqr 
where in this case there are up to a general $n$ number of 
bulk-boundary scalar integral functions.  The translation to 
Schwinger form to compare with a non-holographic field theory 
expression is 
\bqr  
I_{a;a_j}^{\gamma} = C^{n;\gamma}_{a;a_j} \int \prod_{i=1}^n d\alpha_i ~ 
\delta(1-\sum \alpha_j) ~ {\alpha_j^{\Delta_j-1} 
 \over M^{{1\over 2}\sum_{j=1}^4 \Delta_j}}~ M^{\gamma\over 2} \ .
\label{npoint}
\fqr  
The matrix in \rf{npoint} is $M=\sum \alpha_i \alpha_j y_i\cdot y_j$, 
where the vectors ${\vec y}_i$ are chosen to agree with the routing of 
those ${\vec x}_j$ in the integral \rf{npointhol}.
The form in \rf{npoint} is a scalar field theory $n$-point one-loop 
diagram modified as in \rf{npoint} in the general case of 
non-vanishing $\gamma$.  The relation is depicted in Figure 5.  The 
$n$-point correlation functions found from expanding the higher-point 
IIB covariantized string amplitudes have this form, together with 
further $m<n$ point contributions found from multiple $n$ variations 
of the $m$-point covariantized amplitudes of the graviton scattering 
involving terms for example of the form 
\bqr 
S^{n-pt}_{\rm IIB} = \int d^{10}x \sqrt{g}~ \Bigl[ f_{k,g}(\tau,\bar\tau) 
\Box^k R^n + {\rm (2n+2k)-derivative~terms}\Bigr] \ .
\fqr 
We expect the general structure of the one-loop $n$-point field theory 
interpretation of the dual to the IIB superstring scattering to persist 
at finite $\lambda$ and $N$ for this reason in the covariantized 
scattering approach.  

\begin{figure}
\begin{center}
\epsfig{file=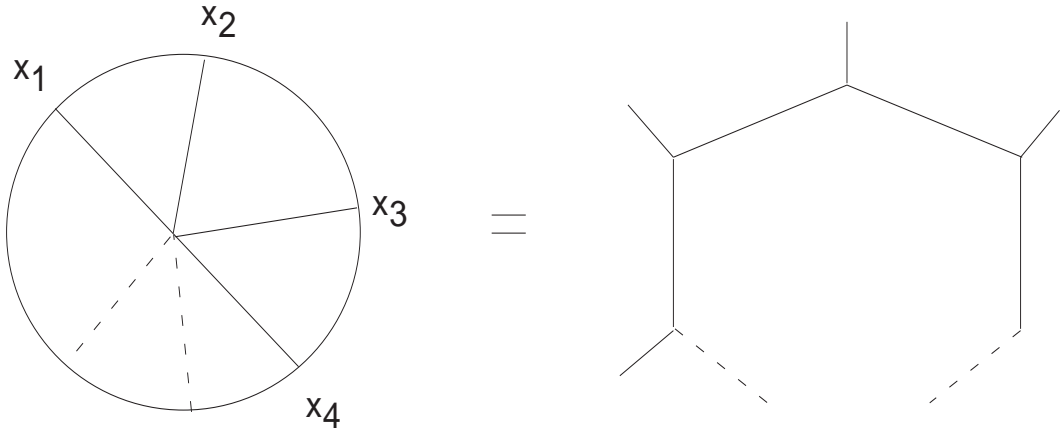,height=6cm,width=14cm} 
\end{center} 
\caption{The $n$-point exchange in the low-energy covariantized 
scattering has the integral form of an $n$-gon integral function 
in $d$ dimensions.} 
\end{figure} 

\section{Discussion} 
\setcounter{equation}{0}  

In this work we have examined the dual relation between IIB superstring 
theory and $\N=4$ super Yang-Mills theory, exploring  a number 
of issues: 1) a formulation of the $\N=4$ gauge theory which is $S$-dual 
to the planar limit, 2) the implications of the former 
for the genus truncation property of the IIB superstring theory at finite 
couplings, 3) the space-time structure of contributions to the correlation 
functions in the gauge theory arising from string theory, 
4) the pole structure of the OPE for these correlators, 5) features associated 
with infinite $N_c$ and unitarity in the correspondence including one-parameter 
integrations associated with correlators in accord with K\"allen-Lehmann 
representations.  

The determination of the space-time integral forms and the coupling 
structure is a step towards finding the operator product expansion 
in $\N=4$ super Yang-Mills theory at intermediate values of the coupling 
constants.  Given the strong form of the AdS/CFT duality and the 
self-equivalence of IIB superstring theory, the OPE structure is related 
to the string scattering at different energy scales.  We have not analyzed 
the non-analytic terms in the string scattering in detail within this 
work, but rather the class of polynomial terms occuring in the 
expansion of the string S-matrix.  These terms generate a class of 
configuration space triangle diagrams on the field theory side 
(generally box diagrams, which due to conformal invariance may reduced 
to triangle diagrams by performing an inversion), whose form suggests 
that these integral functions are dual to the weak-coupling Feynman 
diagrams contributing to the correlation function.  

The one-parameter integral representations in Section 4 of the large 
$N$ limit representing the space-time results on the boundary resemble 
K\"allen-Lehmann forms in perturbative supersymmetric Yang-Mills theory.  By 
virtue of the strong form of the duality, the infinite curvature limit corresponds 
to free-field results for the correlators.  A deduction of Feynman diagram 
representations of the correlators in this limit would demonstrate the 
duality at different regions in coupling space.  The presence of non-renormalization 
theorems indicates that this limit can be taken.  

The constraints of $S$-duality and kinematics imply many features of 
the superstring scattering and the $\N=4$ gauge theory.  In particular, 
S-duality indicates additional cancellations in the truncation of the IIB 
superstring to its massless modes in backgrounds preserving all of the 
supersymmetry.  Via the strong form of the AdS/CFT correspondence, this 
is supported by the existence of an S-dual of the planar expansion in $\N=4$ 
super Yang-Mills theory.   

In the $S$-dual of the 't Hooft limit  there is no  classical supergravity description.  
However in this limit the coupling structure is such that the superstring theory is 
potentially to be described by $\N=4$ topological string calculations \cite{Berkovits:1998ex}.  
This relation between the $S$-dual limit and the $\N=4$ topological string theory 
indicates that there may be a simple description of non-supersymmetric gauge theory in 
this strong-coupling regime, which may be obtained by a resummation of the perturbative 
regime or by a string description.  

\subsection*{Acknowledgements}    
The work of G.C. is supported in part by the U.S. Department of 
Energy, Division of High Energy Physics, Contract W-31-109-ENG-38.  
J.E., who is a DFG Emmy Noether fellow, acknowledges funding through 
a DAAD postdoctoral fellowship.  G.C. thanks David Kutasov for a 
useful discussion.

\vfill\break
\section{Appendix} 
\setcounter{equation}{0}

\subsection{ $t^8$ tensor}

The tensor $t^8$ has the explicit form
\[
t^{8,\mu_1\nu_1 \mu_2\nu_2 \mu_3\nu_3 \mu_4\nu_4} 
 = -{1\over 2}\epsilon^{\mu_1\nu_1 \mu_2\nu_2 \mu_3\nu_3 \mu_4\nu_4} \nonumber 
\]
\[
 -{1\over 2}\Bigl[ 
 (\eta^{\mu_1\mu_2} \eta^{\nu_1\nu_2} - \eta^{\mu_1\nu_2} \eta^{\mu_2\nu_1}) 
 (\eta^{\mu_3\mu_4} \eta^{\nu_3\nu_4} - \eta^{\mu_3\nu_4} \eta^{\mu_4\nu_3})   
 + (2\leftrightarrow 3), (2\leftrightarrow 4) \Bigr] \nonumber
\]
\begin{equation}
+{1\over 2}\Bigl[ \eta^{\nu_1\mu_2} \eta^{\nu_2\mu_3} 
\eta^{\nu_3\mu_4} \eta^{\nu_4\mu_1} 
 + (1\leftrightarrow 2),(2\leftrightarrow 3) 
 + {\rm anti-} (\mu_j\leftrightarrow \nu_j) \Bigr]\,  .
\label{t8tensor} 
\end{equation}
The component of the tensor containing the epsilon tensor does not 
contribute when restricted to the bulk five-dimensional anti-de Sitter space. 

\subsection{Terms necessary for determining the space-time dependence}

\begin{eqnarray}
\lefteqn{\E_{ij,kl}\I^R_{0msklrn0}(z-\x)}\nonumber\\ 
&=&\E_{ij,kl} \left[ -\frac{1}{16}\frac{1}{d^2(d+1)(d-1)(d-2)}
\delta_{mn}\pr^x_k\pr^x_l\pr^x_r\pr^x_s\frac{z_0^d}{(z-\x)^{2(d-2)}} \right.
\nonumber\\
&-&
 \frac{1}{8}\frac{1}{d^2(d-1)(d+1)}\delta_{mn}
\delta_{rs}\pr_k\pr_l\frac{z_0^d}{(z-\x)^{2(d-1)}}
\nonumber\\
&+&
\frac{1}{8}\frac{1}{d^2(d+1)}\delta_{mn}\delta_{rk}\pr_s\pr_l\frac{z_0^d}{(z-\x)^{2(d-1)}}
\nonumber\\
&+&
\frac{1}{2}\frac{1}{d(d+1)}\delta_{mk}\delta_{nl}\pr_r\pr_s\frac{z_0^{(d+2)}}{(z-\x)^{2d}}
\nonumber\\
&-&\left. \frac{1}{4}
\frac{(d-1)}{d(d+1)}
\delta_{mk}\delta_{nl}\delta_{rs}\frac{z_0^d}{(z-\x)^{2d}}
+
\frac{1}{2(d+1)}\delta_{mk}\delta_{nl}\delta_{rs}\frac{z_0^{d+2}}{(z-\x)^{2(d+1)}}
\right]\nonumber \\ &
-& ( m \leftrightarrow s, n \leftrightarrow r) \, , 
\end{eqnarray}

\begin{eqnarray}
\lefteqn{\E_{ij,kl}\I^R_{0msklr00}(z-\x)}&& \nonumber\\ 
&=&\E_{ijkl}\left[
  \frac{(-1)}{2(d+1)}\delta_{lr}\delta_{k[s}\pr^x{}_{m]}\frac{z_0^{d+1}}{(z-\x)^{2d}}
+ \frac{1}{(d+1)}\delta_{lr}\delta_{k[m}\pr^x_{s]}
\frac{z_0^{d+3}}{(z-\x)^{2(d+1)}}\right.\nonumber\\
&+&\left.
\frac{1}{4(d-1)d(d+1)}\delta_{k[m}\pr_{s]}\pr_l\pr_r\frac{z_0^{d+1}}{(z-\x)^{2(d-1)}}
+\frac{1}{2(d+1)}\delta_{k[m}\delta_{s]r}\pr_l 
\frac{z_0^{d+1}}{(z-\x)^{2d}}\right]\quad,
\end{eqnarray}

\begin{eqnarray}
\lefteqn{\E_{ij,kl}\I^R_{00sklr00}(z-\x)} \nonumber\\
&=&\E_{ij,kl}\left[ \frac{1}{16 d^2(d-2)(d+1)}\pr_r^x\pr_s^x\pr_k^x\pr_l^x
\frac{z_0^d}{(z-\x)^{2(d-2)}}\right.\nonumber\\
&-& \frac{1}{8d^2(d-1)(d+1)} 
\left(\delta_{sk}\pr_r\pr_l\frac{z_0^d}{(z-\x)^{2(d-1)}}
+ \delta_{rl}\pr_s\pr_k\frac{z_0^d}{(z-\x)^{2(d-1)}}\right)\nonumber\\
&+&
\frac{1}{4}\frac{1}{d(d+1)}\left(\delta_{sk}\pr_r\pr_l\frac{z_0^{d+2}}{(z-\x)^{2d}}
+\delta_{rl}z_0^{d+2}\pr_s\pr_k\frac{1}{(z-\x)^{2d}}\right)\nonumber\\
&+&
\frac{1}{4}\frac{1}{(d-1)^2
  d(d+1)}\delta_{rk}\delta_{sl}\frac{z_0^d}{(z-\x)^{2d}}
-\frac{d}{(d+1)}\frac{z_0^{d+2}}{(z-\x)^{2(d+1)}}\delta_{rk}\delta_{sl}
+ \frac{z_0^{d+4}}{(z-\x)^{2(d+2)}}\delta_{rk}\delta_{sl}\nonumber\\
&-&\left. \frac{1}{4d^2(d+1)}z_0^{d+2}\delta_{rs}\pr_k\pr_l\frac{1}{(z-\x)^{2d}}
+ \frac{1}{8d^2(d+1)}\delta_{rs}z_0^d\pr_k\pr_l
\frac{1}{(z-\x)^{2(d-1)}}\right] \, .
\end{eqnarray}

\subsection{Box diagrams}

The general box diagram required for the evaluation of the finite $\lambda$ 
and $N$ results is of the form, 
\bqr  
I_{a;a_j}^{\gamma} = \int {dz_0 d^d{\vec z}\over z_0^{d+1}} ~z_0^\gamma 
\times \prod_{j=1}^4 \Bigl[ {z_0^{d+a_j}\over (z-{\vec x}_j)^{2d+2a_j}}\Bigr] \ , 
\label{genboxintegral}
\fqr 
with $\gamma = a+m-\sum_{j=1}^4 a_j$.  The integral form in 
\rf{genboxintegral} is of a box diagram with $d+a_j$ powers of 
propagators on each internal line.  We may express the integral 
in a more conformal field notation via 
\bqr  
I_{a;a_j}^{\gamma} = \int {dz_0 d^d{\vec z}\over z_0^{d+1}} ~z_0^\gamma 
\times \prod_{j=1}^4 \Bigl[ {z_0^{\Delta_j}\over (z-{\vec x}_j)^{2\Delta_j}} 
\Bigr] \ , 
\fqr  
with $\Delta_j=d+a_j$.  The conformal invariance at $\gamma=0$ of the 
integral demands the structure, 
\bqr  
I_{a;a_j}^{0} = F(\zeta,\eta) \times \prod_{i<j}
(x_{ij}^2)^{-{{\Delta_i+\Delta_j}\over 2} +{1\over 6}\sum_{j=1}^4 \Delta_j} 
\ ,  
\fqr 
with the two independent scalar conformal invariant cross ratios in 
\rf{crossratios}.  

The Schwinger parameterization of the integrals in \rf{genboxintegral} 
leads to the integral form, 
\bqr  
I_{a;a_j}^\gamma = C_{a;a_j}^\gamma \int \prod_{j=1}^4 d\alpha_j ~ 
\delta(1-\sum \alpha_i) ~ M^{\gamma\over 2} ~ 
 {\alpha_j^{\Delta_j-1}\over M^{{1\over 2} \sum_{j=1}^4 \Delta_j}} \ , 
\label{schwingerform}
\fqr 
with 
\bqr  
C_{a;a_j}^\gamma = {1\over 2} \Omega_{d+1}^{(a+m)}  
{\Gamma(2d+{a+m\over 2}) \Gamma(2d-{\gamma\over 2}+{1\over 2}\sum a_j) 
\over \prod_{j=1}^4 \Gamma(d+a_j)}  
\fqr 
and 
\bqr  
M = \sum_{i\neq j}^4 \alpha_i\alpha_j x_{ij}^2 \ . 
\fqr 
The $M^{\gamma/2}$ factor represents the difference from the standard 
four-point scalar conformal diagram without any derivatives, i.e. the 
box diagram.  Note that with non-zero $\gamma$ the form in \rf{schwingerform} 
is that of a conformal box diagram evaluated in dimension $d=4-a+\sum a_j$.  

The integrals in \rf{genboxintegral} for $\gamma=0$ are finite; the range 
of $a_j$ is $\vert a_j\vert\leq 2$ and for $d\geq 4$ the arguments of the 
gamma functions are finite.  However, for large enough values of $\gamma$, 
the prefactor in $C_{a;a_j}^\gamma$  develops a pole after regulating, for 
example, with dimensional continuation  ($d=m-\epsilon$ and $m$ integer); 
the origin of this pole is an ultra-violet  divergence found in standard 
field theory by integrating a box diagram with multiple insertions of loop 
momenta in the numerator.  For this reason  the integrals in these cases 
have to be evaluated to first order in $\epsilon$.  

We note that in $d=4$ and for $\gamma=0$ the integrals above may be 
found by multiple differentiation with respect to ${\vec x}_{ij}^2$ of the 
scalar box function 
\bqr  
B= \int \prod d\alpha_j ~{\delta(1-\sum \alpha_j)\over (\mathop{\sum}_{i\neq j} 
\alpha_i \alpha_j {\vec x}_{ij}^2)^2} \ .
\fqr


\medskip 
\begin{thebibliography}{99}  

\bibitem{Maldacena:1998re}
J.~Maldacena,
Adv.\ Theor.\ Math.\ Phys.\  {\bf 2}, 231 (1998)
[hep-th/9711200].

\bibitem{Gubser:1998bc}
S.~S.~Gubser, I.~R.~Klebanov and A.~M.~Polyakov,
Phys.\ Lett.\  {\bf B428}, 105 (1998)
[hep-th/9802109].

\bibitem{Witten:1998qj}
E.~Witten,
Adv.\ Theor.\ Math.\ Phys.\  {\bf 2}, 253 (1998)
[hep-th/9802150].

\bibitem{Aharony:1999ti}
O.~Aharony, S.~S.~Gubser, J.~Maldacena, H.~Ooguri and Y.~Oz,
[hep-th/9905111].

\bibitem{Freedman:1999tz}
D.~Z.~Freedman, S.~D.~Mathur, A.~Matusis and L.~Rastelli,
Nucl.\ Phys.\  {\bf B546}, 96 (1999)
[hep-th/9804058].

\bibitem{Muck}
W.~Muck and K.~S.~Viswanathan,
Phys.\ Rev.\  {\bf D58}, 041901 (1998)
[hep-th/9804035].

\bibitem{Chalmers:1999xr}
G.~Chalmers, H.~Nastase, K.~Schalm and R.~Siebelink,
Nucl.\ Phys.\  {\bf B540}, 247 (1999)
[hep-th/9805105].

\bibitem{Lee:1998bx}
S.~Lee, S.~Minwalla, M.~Rangamani and N.~Seiberg,
Adv.\ Theor.\ Math.\ Phys.\  {\bf 2}, 697 (1998)
[hep-th/9806074].

\bibitem{D'Hoker:1999tz}
E.~D'Hoker, D.~Z.~Freedman and W.~Skiba,
Phys.\ Rev.\  {\bf D59}, 045008 (1999)
[hep-th/9807098].

\bibitem{Skiba:1999im}
W.~Skiba,
Phys.\ Rev.\  {\bf D60}, 105038 (1999)
[hep-th/9907088].

\bibitem{frex}
E.~D'Hoker, D.~Z.~Freedman, S.~D.~Mathur, A.~Matusis and L.~Rastelli,
hep-th/9908160.

\bibitem{SM}
M.~Bianchi and S.~Kovacs,
Phys.\ Lett.\  {\bf B468}, 102 (1999)
[hep-th/9910016].

\bibitem{MJ}
J.~Erdmenger and M.~P\'{e}rez-Victoria,
hep-th/9912250, to appear in Phys.~Rev.~{\bf D}.

\bibitem{MJDE}
E.~D'Hoker, J.~Erdmenger, D.~Z.~Freedman and M.~P\'{e}rez-Victoria,
hep-th/0003218.

\bibitem{HWex}
B.~U.~Eden, P.~S.~Howe, E.~Sokatchev and P.~C.~West,
hep-th/0004102.

\bibitem{Chalmers:2000zg}
G.~Chalmers,
to appear in Nuclear Physics {\bf B}, hep-th/0001190.

\bibitem{Banks:1998nr}
T.~Banks and M.~B.~Green,
JHEP {\bf 9805}, 002 (1998)
[hep-th/9804170].

\bibitem{GG} M.~B.~Green and M.~Gutperle,  Nucl.~Phys.~{\bf B498}, 195 (1997), 
[hep-th/9701093]. 

\bibitem{Intriligator:1999ig}
K.~Intriligator,
Nucl.\ Phys.\  {\bf B551}, 575 (1999)
[hep-th/9811047].

\bibitem{Intriligator:1999ff}
K.~Intriligator and W.~Skiba,
Nucl.\ Phys.\  {\bf B559}, 165 (1999)
[hep-th/9905020].

\bibitem{Brodie}
J.~H.~Brodie and M.~Gutperle,
Phys.\ Lett.\  {\bf B445}, 296 (1999)
[hep-th/9809067].

\bibitem{Berkovits:1998ex}
N.~Berkovits and C.~Vafa,
Nucl.\ Phys.\  {\bf B533} (1998) 181
[hep-th/9803145].


\bibitem{Bsem} 
N.~Berkovits, Proceedings Strings '98, http://www.itp.ucsb.edu/online/strings98. 

\bibitem{Russo:1998fi}
J.~G.~Russo,
Phys.\ Lett.\  {\bf B417}, 253 (1998)
[hep-th/9707241].

\bibitem{Russo:1998vt}
J.~G.~Russo,
Nucl.\ Phys.\  {\bf B535}, 116 (1998)
[hep-th/9802090].  

\bibitem{Chalmers:1999ap}
G.~Chalmers and K.~Schalm,
JHEP {\bf 9910}, 016 (1999)
[hep-th/9909087].

\bibitem{Green:2000pu}
M.~B.~Green, H.~Kwon and P.~Vanhove,
Phys.\ Rev.\  {\bf D61}, 104010 (2000)
[hep-th/9910055].

\bibitem{Sethi} M.~B.~Green and S.~Sethi,
Phys.\ Rev.\  {\bf D59}, 046006 (1999)
[hep-th/9808061].

\bibitem{Obers:2000um}
N.~A.~Obers and B.~Pioline,
Commun.\ Math.\ Phys.\  {\bf 209}, 275 (2000)
[hep-th/9903113].

\bibitem{Obers:1999fb}
N.~A.~Obers and B.~Pioline,
Phys.\ Rept.\  {\bf 318}, 113 (1999)
[hep-th/9809039].

\bibitem{Gopakumar}
R.~Gopakumar and M.~B.~Green,
JHEP {\bf 9912}, 015 (1999)
[hep-th/9908020].

\bibitem{Bianchi}
M.~Bianchi, M.~B.~Green, S.~Kovacs and G.~Rossi,
JHEP {\bf 9808}, 013 (1998)
[hep-th/9807033].

\bibitem{Dorey1}
N.~Dorey, V.~V.~Khoze, M.~P.~Mattis and S.~Vandoren,
Phys.\ Lett.\  {\bf B442}, 145 (1998)
[hep-th/9808157].

\bibitem{Dorey2}
N.~Dorey, T.~J.~Hollowood, V.~V.~Khoze, M.~P.~Mattis and S.~Vandoren,
Nucl.\ Phys.\  {\bf B552}, 88 (1999)
[hep-th/9901128]. 

\bibitem{Belitsky:2000ws}
A.~V.~Belitsky, S.~Vandoren and P.~van Nieuwenhuizen,
hep-th/0004186.

\bibitem{Berkovits:1995vy}
N.~Berkovits and C.~Vafa,
Nucl.\ Phys.\  {\bf B433} (1995) 123
[hep-th/9407190].
 
\bibitem{Eguchi:1998rt}
T.~Eguchi,
hep-th/9804037.

\bibitem{'tHooft:1982tz}
G.~'t Hooft,
Commun.\ Math.\ Phys.\  {\bf 86}, 449 (1982).

\bibitem{Bern:1998ug}
Z.~Bern, L.~Dixon, D.~C.~Dunbar, M.~Perelstein and J.~S.~Rozowsky,
Nucl.\ Phys.\  {\bf B530}, 401 (1998)
[hep-th/9802162].

\bibitem{Liu:1999ty}
H.~Liu and A.~A.~Tseytlin,
Phys.\ Rev.\  {\bf D59}, 086002 (1999)
[hep-th/9807097].

\bibitem{Chalmers:1999wu}
G.~Chalmers and K.~Schalm,
Nucl.\ Phys.\  {\bf B554}, 215 (1999)
[hep-th/9810051].

\bibitem{D'Hoker:1999mz}
E.~D'Hoker and D.~Z.~Freedman,
Nucl.\ Phys.\  {\bf B550}, 261 (1999)
[hep-th/9811257].

\bibitem{D'Hoker:1999jc}
E.~D'Hoker, D.~Z.~Freedman, S.~D.~Mathur, A.~Matusis and L.~Rastelli,
Nucl.\ Phys.\  {\bf B562}, 330 (1999)
[hep-th/9902042].

\bibitem{D'Hoker:1999pj}
E.~D'Hoker, D.~Z.~Freedman, S.~D.~Mathur, A.~Matusis and L.~Rastelli,
Nucl.\ Phys.\  {\bf B562}, 353 (1999)
[hep-th/9903196].

\bibitem{Chalmers:1999xj}
G.~Chalmers,
hep-th/9907015.

\bibitem{Liu:1999th}
H.~Liu,
Phys.\ Rev.\  {\bf D60}, 106005 (1999)
[hep-th/9811152].

\bibitem{Anselmi1}
D.~Anselmi,
Class.\ Quant.\ Grav.\  {\bf 17}, 1383 (2000)
[hep-th/9906167].

\bibitem{Dhoker1}
E.~D'Hoker, S.~D.~Mathur, A.~Matusis and L.~Rastelli,
hep-th/9911222.

\bibitem{Petkou1}
L.~Hoffmann, A.~C.~Petkou and W.~R\"uhl,
hep-th/0002154, hep-th/0002025.

\bibitem{Gelfand}
I.M. Gel'fand and  G.E. Shilov, Ge\-ne\-ralized Functions,
Vol.~I: Properties and Ope\-ra\-tions. Academic Press, New York and London, 1964. 

\bibitem{Osborn:1994cr}
H.~Osborn and A.~C.~Petkou,
Annals Phys.\  {\bf 231}, 311 (1994)
[hep-th/9307010].

\bibitem{Erdmenger:1997yc}
J.~Erdmenger and H.~Osborn,
Nucl.\ Phys.\  {\bf B483}, 431 (1997)
[hep-th/9605009].

\bibitem{Frolov}
G.~Arutyunov and S.~Frolov,
Phys.\ Rev.\  {\bf D60}, 026004 (1999)
[hep-th/9901121].

\bibitem{Tseytlin}
H.~Liu and A.~A.~Tseytlin,
Nucl.\ Phys.\  {\bf B533}, 88 (1998)
[hep-th/9804083].

\bibitem{Tseytlin:2000sf}
A.~A.~Tseytlin,
hep-th/0005072.

\bibitem{Anastasiou:2000ui}
C.~Anastasiou, E.~W.~Glover and C.~Oleari,
Nucl.\ Phys.\  {\bf B572}, 307 (2000)
[hep-ph/9907494].

\bibitem{Boos:1991rg}
E.~E.~Boos and A.~I.~Davydychev,
Theor.\ Math.\ Phys.\  {\bf 89}, 1052 (1991).

\end{thebibliography}
\end{document}